# Ten-tier and Multi-scale Supply Chain Network Analysis of Medical Equipment: Random Failure & Intelligent Attack Analysis[1]

Kayvan Miri Lavassani[a], Zachary M. Boyd[b], Bahar Movahedi[a] and Jason Vasquez[b]

[a]School of Business, North Carolina Central University, Durham, NC, USA; [b]Department of Mathematics, Brigham Young University, Provo, UT, USA

Motivated by the COVID-19 pandemic, this paper explores the supply chain viability of medical equipment, an industry whose supply chain was put under a crucial test during the pandemic. This paper includes an empirical network-level analysis of supplier reachability under Random Failure Experiments (RFE) and Intelligent Attack Experiments (IAE). Specifically, this study investigates the effect of RFE and IAE across multiple tiers and scales. The global supply chain data was mined and analyzed from about 45,000 firms with about 115,000 intertwined relationships spanning across 10 tiers of the backward supply chain of medical equipment. This complex supply chain network was analyzed at four scales, namely: firm, country-industry, industry, and country. A notable contribution of this study is the application of a supply chain tier optimization tool to identify the lowest tier of the supply chain that can provide adequate resolution for the study of the supply chain pattern. We also developed data-driven-tools to identify the thresholds for breakdown and fragmentation of the medical equipment supply chain when faced with random failures or different intelligent attack scenarios. The novel network analysis tools utilized in the study can be applied to the study of supply chain reachability and viability in other industries.

## 1. Introduction

Recent events, such as the COVID-19 pandemic, Suez Canal blockage, Colonial Pipeline cyberattack, Texas ice storm, Huawei ban by the U.S. Department of Commerce, and the 2022 Russian invasion of Ukraine, highlight the pressing need for supply chains that are robust and resilient in the face of high-intensity, sudden disruptions (Nagurney, 2021; Hosseini and Ivanov, 2021; Kosasih and Brintrup, 2021; Crosignani et al., 2021; Powers, 2021; Shi et al., 2022; Boston, 2022).

---
[1]The first and second authors contributed equally.

While the supply chain of many critical products consists of multiple tiers expanded across different industries and countries, relatively little research has focused on the vulnerability issue, with most prior works centering on productivity and efficiency under comparatively small everyday disruptions (Remko, 2020). Indeed, it is not clear how to model large disruptions of the kind observed recently, especially when they affect multiple supply chain tiers across the globe, for which data has been widely considered unavailable. In the present work, we analyze the robustness of the global medical supply chain network, both because of its inherent interest in light of the COVID-19 pandemic and also as a case study to show how such high-intensity, immediate disruptions can be modeled. To do this, we use a novel supply chain discovery technique and utilize an improved metric of supply chain robustness, namely the reachability of terminal suppliers after a network disruption. We conduct our analysis across several scales and using both random and targeted disruptions. Finally, we compare our results to what would have been obtained using classical percolation-theoretical robustness estimates.

While firms traditionally focused on their immediate suppliers to better manage disruptions, Butt, in a 2021 study on the impact of COVID-19 in supply chain networks, recommends that firms should expand the visibility of their supply chain beyond their tier-1 suppliers. The absence of supply chain visibility "creates havoc in the entire value chain" (Pradhan and Routroy, 2008). Supply chain visibility is primarily concerned with extending firms' "timely and accurate" access to "key or useful" information to operate the supply chain (Barratt and Oke, 2007; Barratt and Barratt, 2011; Sunmola, 2021). Despite the importance of supply chain visibility, firms are generally hesitant to share information about their supply chain.

On the one hand, supply chain information sharing among firms is known to result in "improved operational performance, enhanced customer service, reduced costs, improved quality, and enhanced competitiveness" (Yang et al. 2011). On the other hand, many firms have restrictive information disclosure policies that prevent them from promoting visibility across the supply chain (Marshall et al., 2015). There is evidence that justifies the firms' hesitancy in disclosing supply chain information. For example, Mittendorf et al. (2021), in their study of supply chain information sharing, provide evidence that disclosing certain information may not be in the long-term financial interest of the firms or their customers. To address this challenge, we utilize a novel supply chain data collection using data mining of public financial records.

The novel analysis of supply chain viability under failure and attack conditions is another notable contribution of this work that opens the door to fruitful future analytical approaches in the study of intertwined supply chain networks.

The global medical equipment supply chain network can be studied through different scales of network structure spanning across multiple tiers. Motivated by COVID-19, we assess the vulnerability of the medical equipment supply chain at the *network level* (Ivanov and Dolgui, 2021) to various interruptions at the firm, country-industry, industry, and country scales. The present research is primarily concerned with the "supply chain membership" (Marshall et al. 2015) of the firms in the supply chain network.

In this paper, we will discuss how tools from graph theory and statistical physics can be applied to the study of complex supply chain networks. One of the questions that we address in this paper is how many tiers of the supply chain must be analyzed to provide needed visibility to the network pattern. We will also quantify how global supply chains will be affected by different degrees of random and intelligent attacks. Finally, we will provide a methodology to identify the level of failure or attack under which a supply chain breaks down and becomes fragmented.

## 2. Literature Review

Researchers and policymakers have called for development of new solutions for designing and managing global supply chains that are more responsive to "the risk of disruption" (c.f. Sherkarian et al. 2020). The organizational supply chain's ability to manage disruptions is of particular importance "in the time of crisis" for critical supplies such as medical equipment (Okeagu et al., 2021). Okeagu et al. (2021), in their study on the effect of COVID-19 in the U.S. medical system, call for better "transparency of where our raw materials are sourced, diversifying of our product resources, and improving our technology." While companies do not voluntarily report their supply chain information, there are opportunities for data scientists to mine and analyze such data from the available data sources to explore the vulnerability of global supply chains to disruptions. "Disruptions are unexpected events occurring in a supply chain" (Wu et al. 2007) and are closely related to risk and uncertainties in the supply chain (Blackhurst and Wu, 2009). The supply chain's abilities to manage uncertainties and disruptions has been widely studied using various measures, including agility, robustness, vulnerability, flexibility, and adaptability, to name a few. Each measure corresponds to the supply chain's ability to prevent disruption and/or recover from a disruption (Zegordi and Davarzani, 2012). Appendix 1 includes the definition of these abilities.

Supply chain abilities can enable the organization to continue operations despite various uncertainties and disruptions, be they short-term, long-term, minor, or significant. The COVID-19 pandemic is "characterized by a rapid spread" that affected not only supply and demand but also global logistics (Grida et al. 2020). The short-term effect of the COVID-19 interruption has been observed in the daily life of society as well as in critical medical operations (Okeagu et al. 2021). Various export bans of medical equipment that went into effect in 2020 are examples of "protectionism in the pharmaceutical and medical supplies sectors" that put considerable short-term pressure on global supply chains (Stellinger et al. 2020, pp: 23). In the long-term, businesses are expected to adapt to new patterns of production and trade, which stem from operational necessities as well as protectionist policies. "Dismantling the international supply chains, [and] reliance on domestic production" (Yacoub, and El-Zomor, 2020, pp:11) is a real possibility as a result of policies defined by "medical protectionism" and "retreat from global supply chains" (Baldwin and Evenett, 2020, pp: viii) is expected. This study employs a network view of the global supply chain within the general systems theory. We utilize percolation theory and cascading effects to analyze the effect of sources of disruption on the supply chain network.

2.1 *Network View of Supply Chain*
Network analysis is "an essential tool for studying system resilience" due to its capability "to capture relationships and dependencies between components" (Williams and Musolesi, 2016). Advances in data mining and big data computation, along with recent developments in "analysis of...spatial and temporal network[s]," have provided researchers with tools to conduct "more accurate" analysis of many real-world network systems (Williams and Musolesi, 2016). Supply chain optimization practices (Haque et al. 2020) and globalization of markets and production have made the global supply chains less centralized (Abele, Elzenheimer, et al. 2006; Mourtzis and Doukas 2006). Advanced network models have proven capable of analyzing complex networks "even in completely decentralized architectures" (Trajanovski et al. 2012). In their well-cited study on multi-tier supply chain management, Mena et al. (2003) emphasize the network view of the supply chain and describe that contemporary firms are increasingly "operating within more complex and dynamic networks." Mena et al. (2003) explain that not all relationships are of the same importance. For example, one supplier may act as a bridge between different firms or clusters. Thanks to the advancement in the field of network science, we are equipped with several tools to measure the influence of each node in the network and simulate the effect of elimination of each node on the global supply chain network. We employ centrality measures to identify and measure the influence of the firms in the multi-tier supply chain network. In order to be able to conduct such analysis, the supply chain network needs to be mapped across multiple tiers.

2.2 *Supply Chain Mapping*

Studying the global intertwined supply chains often requires mapping the supply chain network (Jia et al. 2019). However, such mapping is "not an effective nor efficient solution" if it is only based on the first-tier suppliers (Choi and Linton 2011). Melnyk et al. (2022) highlight the visibility challenges of multi-tier supply chains and argue that the "only alternative is to undertake some form of supply chain mapping." Mubarik et al. (2022) argue that supply chain mapping can be conducted from three perspectives, "namely, upstream [suppliers] mapping, downstream [customers] mapping, and midstream mapping." However, most firms manage their supply chains by focusing on the first-tier supplier and the first-tier customer; consequently, these firms lack true supply chain visibility. In the present study, we start mapping the supply chain from distributors and use an upstream snowballing approach to map the supply chain. Combining data collection and data analysis methodologies in this paper which are based on network theory and analytical techniques from statistical physics, can be promising for some of the supply chain visibility challenges. These tools can help firms map their supply chains and identify the hot spots within the supply chain quickly and efficiently, as illustrated by mapping and analyzing the medical equipment supply chain in this study.

Furthermore, we identify the most influential nodes at different scales using centrality measures (please see the supplemental data). This is an important contribution since "unless SC mapping is measurable, it is difficult to examine its impact on any performance indicator" (Melnyk et al. 2009). In this paper, we introduce solutions to measure the influence of each node across the global supply chain.

Despite the importance of supply chain mapping in dealing with disruptions such as COVID-19 (Mubarik et al. 2021), constructing such a map is often costly and time-consuming (Hall, 2019). Considering the visibility challenge (Melnyk et al. 2022) of firms' supply chains, some researchers have primarily collected data through case studies on supply chains (c.f. Jia et al. 2019). Collecting data from primary sources is expensive, time-consuming, and may not be feasible to conduct across a large number of firms in an industry on a global scale. The use of primary sources can be helpful for managing the supply of critical components with a relatively few upstream tiers. One example is Toyota's monitoring of its semi-conductor supply chain. Davis ($2021^a$, $2021^b$) explains how Toyota's experience in dealing with the 2011 tsunami in Japan helped the company to be prepared to navigate the 2021 semi-conductor shortage, which was prompted by the COVID-19 and exacerbated by the 2021 fire in Renesas Electronics Corp, one of Toyota's major semi-conductor suppliers. The recent success of Toyota in managing its multi-tier supply chain exhibits the importance of managing the supply chain across multiple tiers.

2.3 *Percolation Theory and Cascading Effect*

The concept of disruption propagation in the supply chain has been well recognized and researched in the area of operations research over the past decades (c.f. Bhamra, Dani, & Burnard, 2011; Ivanov et al., 2013; Ghadge et al. 2021; Sindhwani et al. 2022). The concepts associated with ripple effects, bullwhip effects, and cascading failure, as well as implications of percolation theory in the supply chain, are well studied and understood in the previous research on supply chain disruption. The analysis tools utilized in this study are based on random failures and intelligent (targeted) attacks within percolation theory, as well as the cascading failure of networked structures.

Percolation theory is an extension of the bond percolation process as proposed by Broadbent and Hammersley (1957) to explain how "fluid spreads randomly through a medium" in limestone under gravity using mathematical modeling. Broadbent and Hammersley's (1957) mathematical modeling was an advancement of previous diffusion approaches as they considered the effect of "external forces" to influence the network structure "beside the random mechanism." Percolation theory studies the formation of connected components as nodes or edges are added to a graph (Wierman, 2011), which is naturally dual to the robustness situation, where nodes or edges are removed or disabled. Our work is partly inspired by these ideas, but we notably are not looking at connected graph components; instead are using the reachability of end suppliers, which is more applicable and relevant. It is possible that much of the percolation theory can be extended to this case, but we are not aware of any prior work having done so either theoretically or empirically using real-world big data.

Percolations models have been intensively studied and "applied to a wide range of phenomena in physics, chemistry, biology, and materials science where connectivity and clustering play an important role" (Wierman, 2011). To the best of our knowledge, the earliest application of percolation theory in business was the use of this theory to explain the "spread of innovation" (Hausman, 1998) along with diffusion theory. Both of these theories originate from "physical science and engineering" (Hausman, 1998). Percolation theory found its way to other areas of organizational studies and business topics in the early 2000s through its application in bankruptcy (Aleksiejuk and Holyst, 2001), innovation management (Silverberg and Verspagen, 2003), disaster management (Helbing and Kuhnert, 2003) and in the retail network (Cliquet and Guillo, 2013). Only in the past decade have scholars in the field of operations and supply chain management started to explore supply chain networks' reachability through the lens of percolation. For example, Mizgier, Wagner, and Holyst (2012), in their study on supply chain networks, highlighted the nonlinear nature of supply chain structures due to random failures in the context of bankruptcies using simulated models.

Supply chain failures are usually caused by random events (such as natural disasters, including pandemics) or targeted/strategic events (such as trade wars, terrorism, sanctions, trade regulations, and competitive pressures). Percolation theory provides analytical tools to explore both types of causes of interruption. The study of targeted attacks can also reveal firms' strategic supply chain weaknesses and strengths. A few previous studies have attempted to study the application of percolation theory in the supply chain using simulation or modeling without tracking the multi-tier supply chain to end suppliers. For example, Zhou and Wang (2018) proposed measuring supply chain efficiency[2] under attack scenarios and failure scenarios of nodes and edges using a simulated scale-free model generated by Barabási and Albert's algorithm (Barabási and Albert, 1999). In another study, Viljoen and Joubert (2018) explored the role of transportation infrastructure on supply chain network resiliency using simulation. They argued that most of the previous work around the design of the supply chain had been concerned with quantifying and optimizing the supply chain efficiency, and less attention has been paid to quantifying the supply chain vulnerability. Viljoen and Joubert (2018) promote the use of percolation theory and cascading failure as "preferred" tools to quantify the supply chain network reach and vulnerability assessment.

---

[2] Zhou and Wang (2011) defined supply chain efficiency "as the average of the reciprocal of the shortest path lengths between each node pair in the network".

2.4 *Research Gaps and Cascading Effects*

Mizgier et al. (2012) further call for the need to use "real-world data" in this area. We address this gap in this study. More recent studies have recommended the use of percolation and propagation of cascading effects in risk management, particularly in the area of critical infrastructure and supply chain (c.f. Schauer et al. 2018, Schauer 2018, Smith et al. 2019, Abdulla and Birgisson, 2020, and Wang et al. 2021). Percolation is among the tools that can help firms navigate more successfully in "a post-COVID-19 world" (Hill et al. 2021). Despite many theoretical modeling studies and a few case studies, there exists a gap in the literature on empirically analyzing the complex network topology of the global supply chains through real-world data (Krause et al. 2016).

Addressing this gap requires the application of new methodologies. For example, we need to develop a data-driven methodology to identify the number of supply chain tiers that must be observed to have proper visibility of the supply chain pattern that crosses industries and borders. We discuss our tier-count tool that identifies the minimum number of tiers required for analysis under each disruption scenario. Another novel analysis that we apply from the field of statistical physics is with regard to the identification of a threshold of supply chain breakdown and fragmentation. This addresses another gap in the literature with regard to the threshold of supply chain disruption (c.f. Wang et al. 2018).

The analytical methods used in this study are partially inspired by the cascading approach (c.f. Sauer and Seuring, 2019). Supply chain cascading analysis has been widely used to analyze disruptions in the context of computer network architecture (c.f. Potts et al. 2020), computer network security (c.f. Yan et al. 2014), infrastructure networks such as power-grids (c.f. Ash and Newth, 2007; Guo et al. 2019), and traffic networks (c.f. Li et al. 2019). Cascading failure in these studies is usually caused by the overload of the network, which is not frequently applied to supply chain failure analysis. Yang et al. (2021) argue that "when a node is disrupted, its downstream and upstream neighbors will be affected due to supply shortage and demand losses, respectively." From this perspective, both underload and overload can negatively affect the supply chain robustness through not only inventory but also cost (Sun et al., 2020). For example, in the case of an underload, the supply chain may be disrupted due to unfavorable economies of scale in backward or forward tiers of the supply chain.

As the supply chain of medical equipment has implications in the nations' quality of life and national security, previous studies on infrastructure and military supply resiliency (c.f. Brown et al. 2005, Barrow 2019) have contributed to a better understanding of critical supply chains. Brown et al. (2005) argue that while commercial supply chains may not be generally considered as critical infrastructure, "they are certainly critical" to the "well-being" of a nation. Disturbances to the networked infrastructure may be caused by "random failure, deliberate attacks, and natural disasters" (Wang et al., 2013).

## 3. Research Methodology

In their well-cited paper on supply chain disruption, Wu et al. (2007) discuss the intertwined global supply chain of products and services. Network-based modeling and analysis have been a recommended "methodology for supply chain distortion analysis" that addresses the complex, multi-tier, nonlinear, global, and dynamic characteristics of organizational supply chains (Wu et al. 2007). In arguably one of the most targeted and comprehensive investigations in the supply chain literature about "disruption propagation and structural dynamics," Ivanov and Dolgui (2021) argue that this area has been explored using three categories of methodological tools, namely, network and complexity theory, mathematical optimization, and simulation studies. The present work can be best classified as a network and complexity theory application, where supply chain analysis is conducted on the "macro view of the supply chain structure" and "operational parameters" are not the subject of the research (Ivanov and Dolgui, 2021).

We now discuss some of the network-based methods to study the sources of supply chain disruptions along with their application in the present study. Petri nets have been discussed in the literature as a method for analyzing disruption risk and uncertainty in complex global supply chain networks. While previous works aimed to understand supply chain disruptions using risk probability, Petri-net modeling does not require the availability of such probability distributions, which are usually constructed based on past experiences (Zegordi and Davarzani, 2012). This is good, since it allows for the modeling of events, such as the COVID-19 pandemic, for which we do not have probability estimates. Petri-net models have been utilized by firms with access to detailed product information, including the bill of materials and production process of their products/services. While such information is not available at the global supply chain scale, we adopt the "reachability" analysis approach used in Petri-net modeling across the network (Wu et al., 2007, Zegordi and Davarzani, 2012, Fierro and Garcia, 2020). "Reachability" is the basis of our supply chain disruption analysis. In this paper, we explore how the firms' access (or reach) to their multi-tier supply chain is affected when interruptions occur.

According to percolation theory, in certain classes of networks, there is a threshold number of edges that must be present for the network to form a giant connected component, but below that threshold, the network consists of disconnected clusters (Albert and Barabasi, 2002).

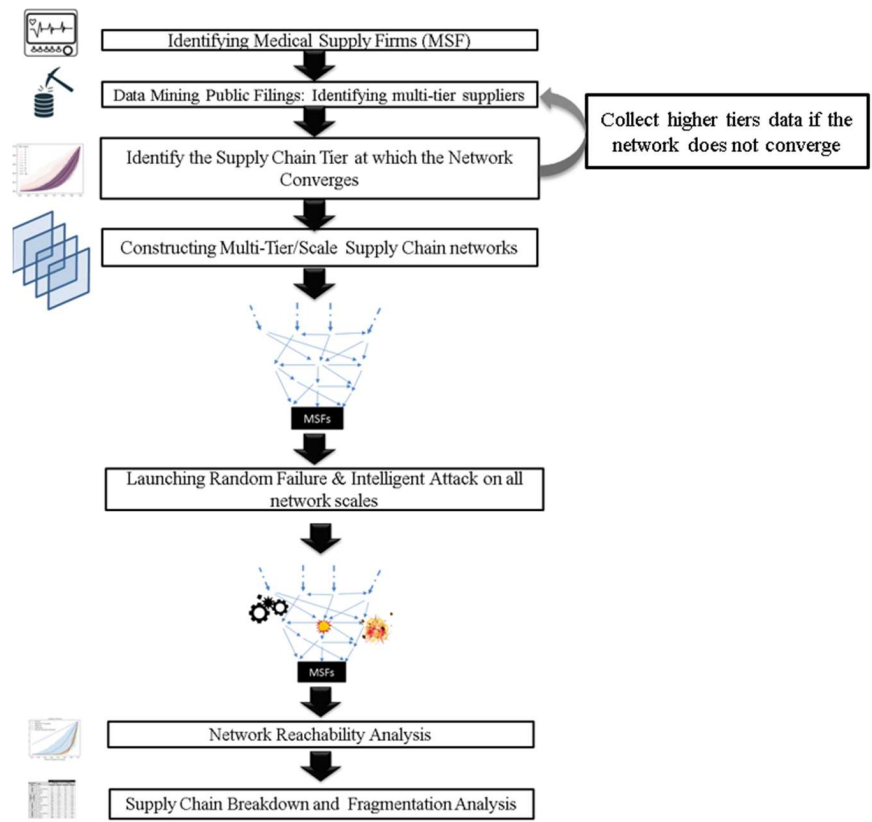

Figure 1 Caption: Data Collection and Analysis

Figure 1 Alt Text: Data collection, preparation, and analysis. After identifying the main suppliers of medical equipment, we conducted data mining to collect and prepare the supply chain network across different tiers. The data is mined from the public financial filings, as detailed in the data collection section.

Figure 1 displays the main steps of data collection and analysis in this work. To ensure that we have collected data from enough backward tiers of supply chain to have a stable resolution of the network, we conducted a convergence analysis. We then prepared data across tiers and scales of analysis. We conducted network reachability analysis using the result of random failure and intelligent attack failures. Finally, identified the threshold of supply chain breakdown and fragmentation at different limits. These steps are described in the data collection and data analysis sections.

There have been no previous studies using the supply chain reachability method on real-world data across large-scale multi-tier supply chain data. The limited number of previous studies that utilized similar methods have explored other aspects of the supply chain, such as supply chain sustainability analysis at small scales or using stochastic methods (c.f. Kumar and Rahman, 2017; Bommel, 2010). In this study, the global supply chain reachability will be assessed in the presence of random failure and intelligent attacks using real-world supply chain data.

## 4. Data Collection

The supply chain data on which this study relies consists of 115,118 real relationships between 44,927 firms, with other scales of the network being computed from the firm-scale data. To the best of our knowledge, the previous studies in this area have been conducted either using stochastic, synthetic (simulated) supply chain networks (c.f. Sen et al. 2020; Yang et al. 2021; Wang et al. 2018) or using relatively small local supply chain structures (c.f.; Hernandez and Pedroza-Gutierrez, 2019, research on seafood market in Guadalajara, Mexico based on the study of 10 wholesalers, using a single tier). Considering that "synthetic networks" are constructed using "random network models" to represent a simulation of multi-tier supply chains (Yang et al. 2021), they can provide an opportunity for researchers to practice various scenarios in synthetic supply chain structures, assuming access to information about the supply chain. Despite the convenience of using simulated networks to study real-world constructs, these simulated networks come with restrictive assumptions and simplifications that limit the implications of the findings. For example, in an innovative study on supply chain resilience and restoration after a crisis using simulation data, Mao et al. (2020), assume that all firms in the supply chain will select the shortest path during restoration, which is not widely applicable to real-world supply chains. Additionally, data scientists using simulated networks face various limitations, including credibility, scalability (Rampfl, 2013), reliability, accuracy, different algorithms' characteristics, and generalizability of the findings (Cassens et al., 2005). The alternative to simulated data has traditionally been carefully curated, real-world data. The advantage of this approach is that the topology is more realistic than in the random graph models, so the results may be more relevant. However, it is very difficult to collect enough real-world data to capture the multi-tier, multi-scale complexity of global supply chains, so in general, studies using real networks can suffer from serious missing data/boundary effects as well as a lack of understanding of how generalizable the conclusions are to broad industries. The present study utilizes a novel data collection method from financial resources to get a comprehensive picture of the entire real-world global supply chain among public firms.

The Securities and Exchange Commission (SEC) Statement of Financial Accounting Standards (SFAS) requires publicly traded firms (and many private firms) to report their notable customers and suppliers, among other information. More specifically, 17 CFR 229.101 requires firms to report their business description along with information about the suppliers and customers that "accounted for 10 percent or more of consolidated" revenue/cost "in any of the last three fiscal years, or if total revenue did not exceed $50,000,000 during any of those three fiscal years, 15 percent or more of consolidated revenue" or cost (SEC, 33-7620). We refer to these firms as "notable" customers and suppliers. These legal requirements, along with similar requirements by regulatory agencies of all other major stock exchanges, provide data scientists access to global supply chain data far beyond what is used in previous supply chain studies. In particular, there is every indication that this data is not biased toward US firms. To the best of our knowledge, this is the most comprehensive large-scale feasible, and legally available supply chain data. The data for this research is prepared and provided by the Innovation &Entrepreneurship Business Ecosystem Lab (IEBE Lab). IEBE Lab uses various data mining techniques and resources to mine supply chain data. This data is not limited to publicly traded firms. Many medium and large private companies who wish to have access to certain financial instruments (e.g., those issuing more than

500 common shares with assets in excess of $10 million) are also required to file their financial information with their exchange commission.

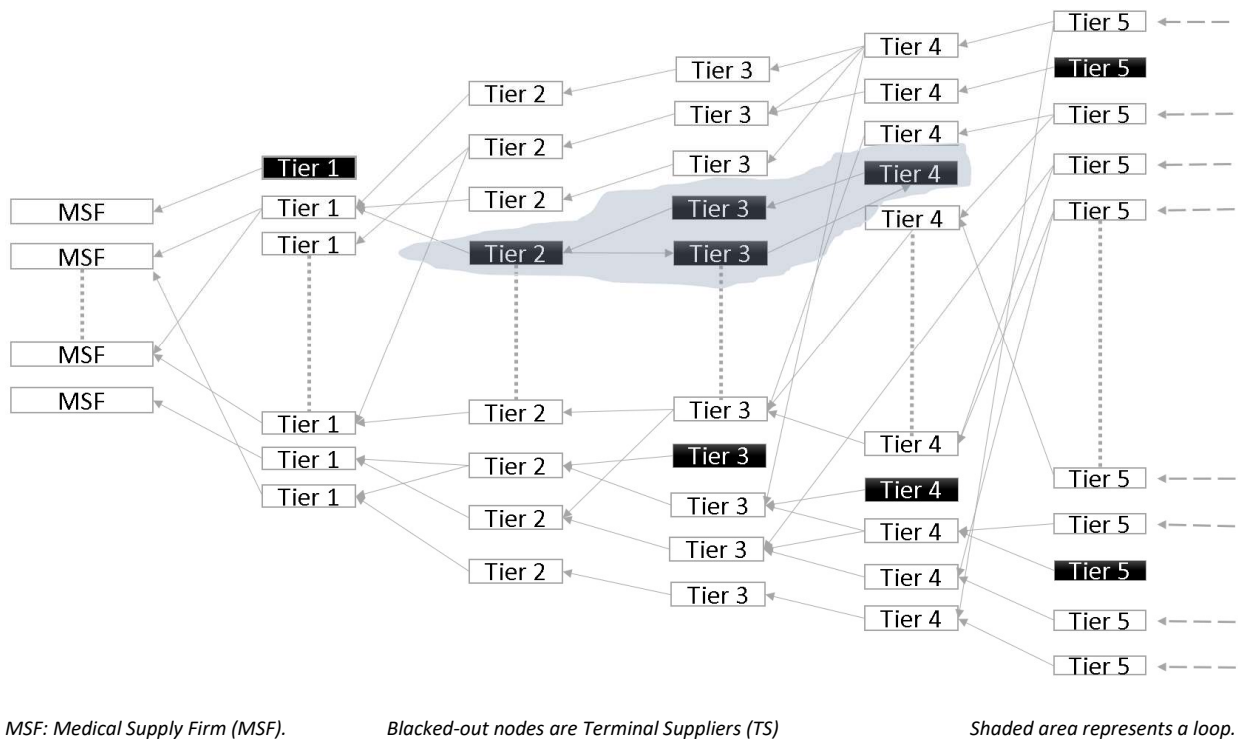

MSF: Medical Supply Firm (MSF).    Blacked-out nodes are Terminal Suppliers (TS)    Shaded area represents a loop.

Figure 2 Caption: Network Data Structure: Tier-by-Tier Data Collection.

Figure 2 Alt Text: An illustration of a multi-tier supply chain across five tiers. The image identifies terminal suppliers. An example of a loop is also displayed in the image.

A schematic map of the first five tiers of this data is presented in Figure 2. As a simplification, Figure 2 presents each tier as disjoint. In reality, our supply chain network is nonlinear, and many firms are present across multiple tiers. Medical Supply Firms (MSFs), which are wholesalers or distributors of medical equipment, are the starting point of data collection. Terminal Suppliers (TS) are firms in the supply chain for which we don't know of any higher-tier dependencies, and they are identified in Figure 2 by black-filled nodes. These TSs are not necessarily in the last tier, as lower-tier firms are not guaranteed to report any notable suppliers in higher tiers. The supply chain network includes numerous cycles. For cycles that include a TS, all members of the cycle are identified as TSs. In Section 5, we have provided the analytical justification for our choice to collect ten tiers of data.

The data collection starts with all companies listed in SNL Financial, S&P Capital I.Q., and Compustat under the primary Standard Industry Classification (SIC) 5047 (Medical, Dental, and Hospital Equipment and Supplies), which was 324 firms at the time of data collection. The supply chain information about notable suppliers of 267 MSFs was available to be mined. Ten rounds of data mining and preparation were performed to construct the 10-tier supply chain network. The network was constructed from 115,118 relationships between 44,927 firms. Appendix 2 illustrates the distribution of the suppliers as well as the number of new firms we mined at each round of data collection. Information about the type of ownership is available for all of the firms in the sample. Twenty-four percent of organizations in our sample are public companies, 73% are private companies, and there is a small number of educational institutions, foundations/charitable institutions, and government institutions, which account for 3% of our sample population. Information regarding the number of employees is available for 63% of the firms in our sample. The median firm size in our sample, as measured by the number of employees, is 463, with 48% of firms in our sample having more than 500 employees. Fifty-two percent of the firms have 500 or fewer employees. A smaller portion of firms in our sample (25%) have 100 or fewer employees. If we assume that all firms with missing employees' data are private Small and Medium Enterprises (SMEs), the share of firms with 500 or fewer employees increases to 69%. Appendix 3 presents the list of 20 main suppliers at each of the four scales of analysis. We assigned each firm to a country-industry using the country where its headquarters is located, following Lavassani's (2017) proposed multi-scale network analysis approach.

## 5. Data Analysis

As sources of disruptions may be at the firm scale, country scale, country-industry scale, or industry scale, it is imperative that businesses and policymakers have the capability to understand and analyze disruptions at all scales. For example, consider global sanctions placed on a particular industry in a specific country. In this case, the country-industry scale would be suitable.

We determined the number of tiers required for our analysis by requiring uniform convergence of the reachability curves (defined below) to within a 5% tolerance. We remark that uniform convergence of relevant statistics is a general-purpose tier count optimization tool that can identify the most efficient supply chain network depth to analyze. Uniform convergence criteria are widely applied throughout the simulation sciences, particularly mathematical physics. (c.f., Scott 2011, chs 12-13 for introductory material.)

In this study, we utilize different random failure and intelligent attack methodologies, which are commonly used to study complex networks in the fields of mathematical physics, and system resiliency (c.f. Liu et al. 2005; Magnien et al. 2011; Yamashita et al. 2019; Sičanica and Vujaklija, 2020). To analyze the effect of disruptions, we conducted Random Failure Experiments (RFEs) and Intelligent Attack Experiments (IAEs) on the global supply chain across different tiers and scales. The RFEs and employee-based IAEs are performed using 100 realizations. Due to missing industry categorization for some firms, we also repeat the industry-level PageRank-based experiments 24 times, each with a different imputation of the missing industry values, drawn with replacement from the distribution of industries that are known. We also plot the percentile intervals of outcomes from random and randomized attacks, from the $2.5^{th}$ percentile to the $97.5^{th}$ percentile. PageRank, as a centrality measure in our static observed graph, is an observed variable as it can be measured directly without inference from other observed variables. Centralities were calculated at the firm-scale. The centralities at larger scales are taken to be the sum of constituent firm-scale centralities.

Unlike random network models such as Ising models, our static network does not bear a possibility of illustrating different structures, and hence its properties, such as PageRank centrality, do not directly qualify as latent variables (Hallquist et al., 2019). Similar to other observed variables, should there exist a theoretical support, a centrality measure can be potentially combined with other measures through factor analysis techniques within psychometric methodologies to create latent variables (c.f. Hallquist et al. 2019; Joshanloo, 2021). The variables used in this study are observed variables, which are directly calculated based on the observed dyadic multi-tier network of the global supply chain.

Our procedure includes a set of procedural and conceptual assumptions that are presented in Table 1 and are briefly discussed in the following.

Table 1. Operating Assumptions

| **Operating Assumptions** | |
|---|---|
| ● New substitute relationships are not allowed | ● No capacity limit |
| ● Alternate existing path feasible | ● Time lag |
| ● Unweighted network | ● Limited to the network of notable suppliers |

The no substitution assumption assumes that if a supplier is eliminated, the node (firm, industry, industry-country, or country) cannot substitute that supplier with another supplier through establishing a new path. The "alternate existing path feasible" assumption implies that while the affected node cannot establish a new path, the node is still allowed to use the existing alternate path or paths that may already exist in the network.

The "no substitution" assumption may not be very realistic at the firm level since firms can switch suppliers, albeit after incurring switching costs and delays. However, the no substitution assumption can be materialized at the higher scales. For example, the substitution may not be feasible at the industry level due to the product specification. The "alternate path" assumption, while possible, may not be feasible, especially at the firm scale, due to a partial or complete lack of compatibility of suppliers that may limit product substitutability and due to available capacity throughout the alternate path.

The procedure also assumes that all edges have equal weight in the network. While we know that the supply chain relationships are notable, we do not have data about the dollar value of the transactions. Also, when a node is affected, and alternate network edges are utilized, we are assuming that the existing nodes and edges have the capability to satisfy the additional demand.

The supply chain network includes many large firms, and SMEs that do not have substantial financial transactions with large firms may not be identified in our data mining process.

Although the supply chain data collected from annual filings are accurate as of the date of data collection, and it is expected that companies do not change their notable suppliers very frequently, still, the network can include firms that are no longer notable suppliers. The time lag is more limiting in the interpretation of firm scale data.

For conducting the IAEs, we need to define target attack criteria. We selected PageRank centrality as a recommended measure of "attack-based resiliency analysis" (Zhao et al., 2015; Newman, 2006). We use PageRank on the network and its transpose as two reasonable proxies for importance with respect to upstream and downstream firms, respectively. High-PageRank firms were targeted first in the PageRank IAEs. We also looked for other available moderating factors that could affect firms' influence in the supply chain network. We could collect the number of employees for 95% of firms in our supply chain network. In the absence of edge weights, the number of employees is a meaningful measure of a firm's size and influence in the supply chain network. Thus, in the employee-based IAEs, the firms with the most employees were calculated first. (For firms without employee data, employee counts were imputed similarly to the industry imputation described above.) To measure the consequence of RFE and IAE on the global supply chain, we measured the average percentage of TSs Reachable (ATSR) as well as whether at least one TS was reachable (Some Terminal Suppliers Reachable, or STSR), averaged across MSFs.

Finally, we present two data-driven approaches to identify the threshold of disruption that may lead to the breakdown and fragmentation of the global supply chain for medical equipment. The data analysis is presented in six subsections, namely, supply chain tier, firm scale, industry scale, country-industry scale, country scale, and supply chain breakdown and fragmentation.

*5.1 Supply Chain Tiers Analysis*

The number of tiers required for these analyses was data-driven, using uniform convergence of reachability statistics as a stopping criterion. The supply chain data is collected tier by tier, and the data can be collected virtually across an unlimited number of tiers. However, due to the large size of the data and computational limitations, a data scientist should identify the optimal number of tiers that best represent the structure of the global supply chain pattern. This issue is crucial since, as we expand data collection from one tier to the next, the number of firms in each supply chain tier may grow exponentially. Our analysis across different scales provides mathematical evidence for the appropriate number of tiers that can best represent the supply chain pattern for each type of analysis (i.e., random failures vs. intelligent attacks) and each scale of analysis (e.g., firm- vs. country-scale). Figure 3 displays the result of RFE and IAE across the four scales of analysis.

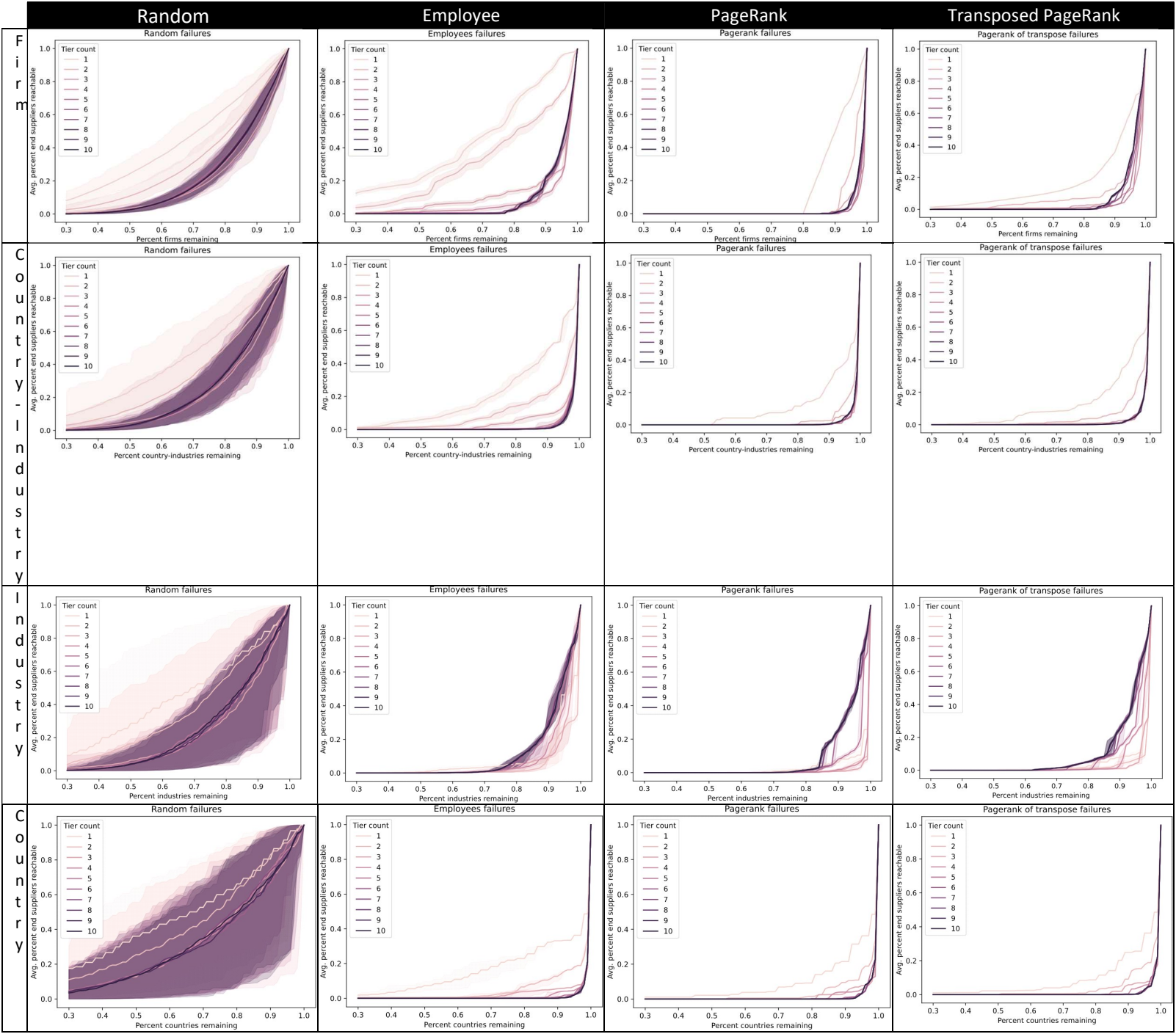

X: One minus the Supply Chain Failure Rate (SCFR)
Y: Average Percent of Terminal Suppliers Reachable (ATSR)
In the random/randomized failure analysis: Realizations= 24 or 100.
Shading represents the range from the bottom 2.5th percentile to the top 97.5th percentile interval over realizations.

Figure 3 Caption: RFE & IAE Across Multiple Tiers of Supply Chain.

Figure 3 Alt Text: A sixteen-panel image of RFE & IAE at four scales across ten supply chain tiers.

The Y-axes in Figure 3 represent the average percentage of the end suppliers which are reachable (where the average is taken across MSFs); this statistic is equal to one minus the Supply Chain Failure Rate (SCFR). The X-axes indicate the percent of remaining operating firms, representing the percentage of firms that are still able to produce their goods or services under the RFE and IAE.

Our analysis demonstrates that the SCFRs across different tiers in the medical equipment industry have high correlations. The shaded areas capture 95% of the data spread across random realizations of firm failure order and/or industry/employee count imputation where needed. This shaded area shows the range of probable disruption variation when the supply chain experiences failure. Each of the RFEs or IAEs in Figure 3 allows for a visual assessment of network convergence. To have a reasonably appropriate estimation of the supply chain structure, researchers need to collect data from an appropriate number of tiers. To identify the tier in which the network converges, we use uniform convergence of the mean as a function of percent firms remaining. We subtracted the mean ATSR curve with 10 tiers from the one with fewer tiers and take the absolute value. The largest value of the resulting function is then the uniform distance. Starting with one tier and adding tiers until the uniform distance is small enough (5% in our tests) we identified how many tiers are needed for convergence. Based on the analysis of each source of disruption across different scales of supply chain, we recommend a minimum number of supply chain tiers to be analyzed for each disruption scenario, as displayed in Table 2.

Table 2. Disruption scena

| Failure/Attack mode \ Unit | RFE Random | RFE PageRank of transpose | IAE PageRank | IAE Employee |
|---|---|---|---|---|
| Firm | 4 | 8 | 7 | 6 |
| Country-Industry | 3 | 5 | 6 | 4 |
| Industry | 5 | 8 | 8 | 7 |
| Country | 4 | 4 | 7 | 6 |

According to our analysis, for the purpose of RFE analysis at the firm-scale, we need to collect data from at least four supply chain tiers. In this scenario, the supply chain's reachability pattern stabilizes after the fourth tier; we call this the convergence tier count. Including higher tiers of the supply chain in this scenario only marginally enhances the network's resolution of the supply chain pattern, while substantially increasing the data complexity. Another example is the IAE based on employee count at the firm scale, where six tiers provide evidence that required supply chain information is collected to have a converged supply chain pattern for further research.

The decision about "data acquisition" is a critical one as it raises the question about how many more tiers are "economically reasonable" (Mohr and Rijn, 2022). Additionally, larger data carries the need for a more sophisticated data processing machine and more efficient algorithms. This critical decision has been of interest in the area of data science generally, with applications in various areas such as object recognition (c.f. Zhu et al. 2016), machine learning (c.f. Pedregosa et al., 2011, Mohr and Rijn, 2022, and Cui et al. 2022), and physics simulation validation (c.f. Zhao and Su, 2019). In general, to enhance the accuracy of a model, we can either enhance our algorithms or enhance the data quantity/quality—ideally both (Pedregosa et al., 2011; Zhu et al., 2016). As we are working with real data and a defined objective rooted in percolation theory, an effective method to enhance the validity of the model is to collect data from higher tiers and test the convergence of the models. In the present study, our data collection was simultaneous with the model development, so we opted to be as inclusive as possible by capturing essentially all available data. As we collected data from higher tiers, the number of new firms increased up to the $6^{th}$ tier. The number of new firms in the sample decreased beyond the $6^{th}$ tier, and we only had two new observations (firms) added to the supply chain network in the $10^{th}$ tier, which resulted in the termination of further learning (see Appendix 2). Based on the convergence criteria we later developed (uniform convergence), we could have stopped after 8 tiers without significantly affecting the study's accuracy.

We also identified several interesting patterns of supply chain disruptions. One of the interesting observations is related to the case of industry-scale PageRank IAE. In this case, we recommend using at least eight tiers to obtain the best supply chain failure pattern depending on the scale of attack. Six to seven tiers in this scenario would be sufficient for attacks that take out up to approximately 10% of industries; however, for attacks that affect more than the 10% threshold, our analysis suggests collecting data from at least eight tiers. Based on the pattern of failure at this scale of analysis (Figure 3), it is expected that should data be collected from higher tiers, the supply chain may show more robustness and continue to show a stepped-down pattern of catastrophic failure when the percentage of "industries remaining" continues to decrease beyond

0.8. In other words, an analysis with fewer tiers may overemphasize how fragile the supply chain is to this type of targeted attack. Another interesting finding based on the analysis of the pattern of supply chain disruptions across different tiers can be observed in the result of RFE at the country scale and the industry scale. The dark-shaded areas representing the bottom $2.5^{th}$ percentile shown below the mean reachability curves display the possibility of a catastrophic disruption with a relatively small elimination of units. For example, a 5% disruption at the industry scale or at the country scale can potentially cause 80% of MSFs TSs to become unreachable. This is an indication of high cascading dependency of interdependent industry networks where a "small fraction of fault nodes may lead to complete fragmentation of a system" (Hong et al. 2015)

If we consider 20% reachability of TSs from MSFs as a "catastrophic" supply chain failure, we expect this threshold would probably not be reached unless at least 30% of industries or countries are randomly eliminated from the global supply chain of medical equipment. However, our IAE analyses indicate that should even a few (approximately 5%) of the notable industries or countries be eliminated, we can reach a catastrophic supply chain failure. The RFE analysis at the industry scale and the country scale reveals another interesting characteristic: as we collect and analyze supply chain data from higher tiers, the likelihood of higher TSs' reachability (shaded area above the curves) decreases, however, the likelihood of catastrophic failure (shaded area under the curves) does not decrease notably.

We now have supportive evidence that the data includes enough tiers to conduct detailed analyses at all scales. We will discuss these analyses in the following sections.

## 5.2 Firm-scale Supply Chain: RFE and IAE Analysis

Figure 4 displays the RFE and IAE analysis at the firm-scale. To better focus on the data ranges where notable changes occur, the analysis is presented with 0.3-1.0 and 0.9-1.0 remaining firms ranges. We have plotted the ATSR and STSR. Conducting analysis using both ATSR and STSR provides further insight for business strategists. In scenarios where (1) MSFs have buffer inventory for some inputs (e.g., parts), (2) there exist substitute inputs, or (3) MSFs can source some inputs from other suppliers, the STSR can be a better measure of supply chain operability.

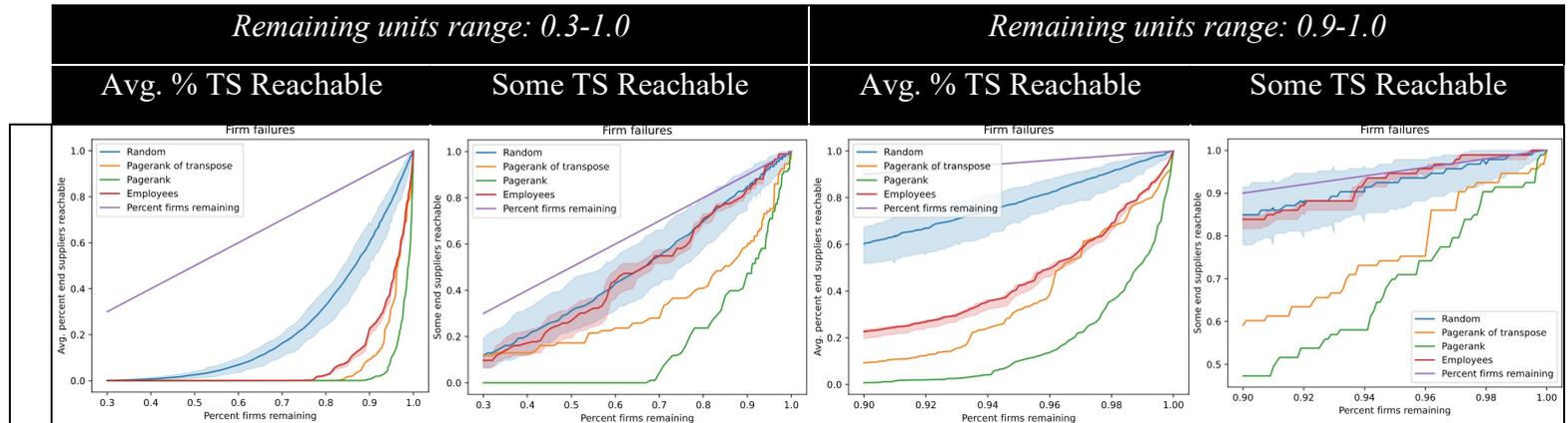

Figure 4 Caption: Firm-scale RFE & IAE of the global medical equipment supply chain.
Figure 4 Alt Text: A four-panel image of RFE & IAE at the firm-scale displays disruption's effect on the supply chain.

Overall, we have supportive evidence that the most effective disruption can be caused by an intelligent attack which is based on the centrality of the firms. Intelligent attacks targeting lower-tier central suppliers (PageRank) are found to be more effective in disrupting supply chains than targeting higher-tier central suppliers (PageRank of the transposed network). Random failures are found to be the least effective in disrupting the supply chain. Finally, intelligent attacks based on firm size (as measured by the number of employees) are found to be similar to random failure on STSR.

It is noteworthy that the vertical distance between the percentage of firms remaining (purple line) and ATSR/STSR is the network effect stemming from supply chain dependencies. For example, in the STSR graphs (Figure 4), we can observe that the network effects of random failure and intelligent attacks are modest in the >95% firms remaining range; however, failures that affect more than 10% of firms will cause notably larger disruptions. In the case of ATSR, the network effects of disruptions are larger and earlier.

### 5.3 Country-Industry-scale Supply Chain: RFE and IAE Analysis

Figure 5 presents the analyses at the country-industry scale. In the ATSR analyses, different intelligent attack methods produce very similar scales of disruptions, even in the 0.9-1.0 range. Similarities can be observed in STSR analyses as well; however, the PageRank-based attacks and transposed PageRank attacks are marginally more effective than size-based attacks.

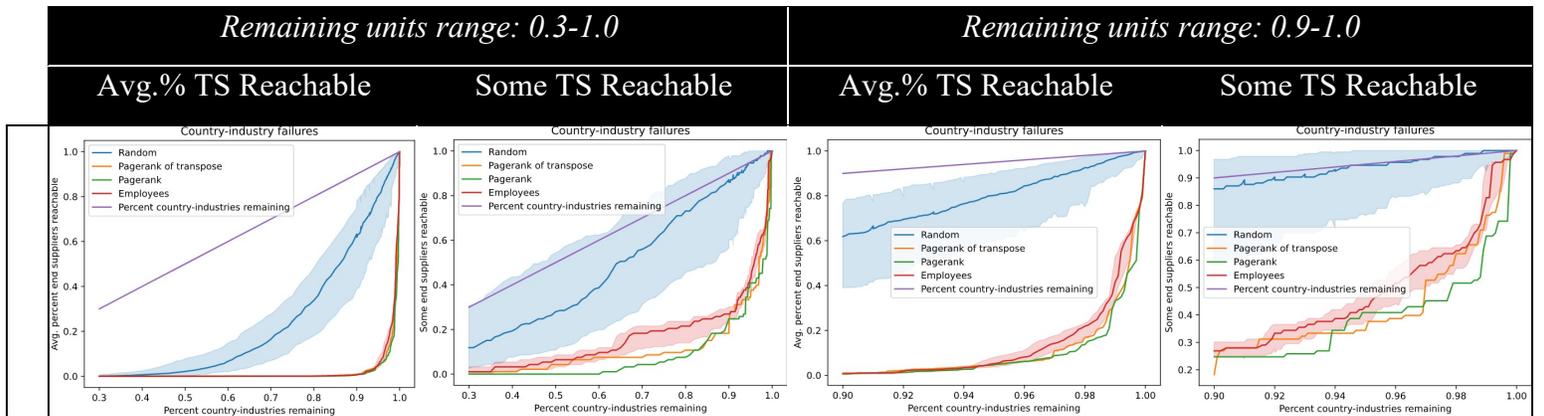

Figure 5 Caption: Country-Industry-scale RFE & IAE of the global medical equipment supply chain.
Figure 5 Alt Text: A four-panel image of RFE & IAE at the country-industry-scale displays disruption's effect on the supply chain.

## 5.4 Industry-scale Supply Chain: RFE and IAE Analysis

The industry-scale analysis (Figure 6) measures the impact of random or targeted elimination of an industry. The two centrality-based attacks produce nearly identical effects on ATSR and STSR. This implies that attacks targeting higher tiers or lower tiers at the industry-scale are expected to result in the same scale of interruptions. It is notable to mention that the supply chain shows notable resiliency in centrality-based industry attacks under STSR scenarios within a certain attack range. Specifically, we can expect over 80% of TSs to stay reachable when approximately 15% of the industries are eliminated. In this scenario, the random attack has a reasonable probability of causing more devastating supply chain disruption as identified by the shaded area reaching below intelligent attack curves (for certain ranges of remaining industries).

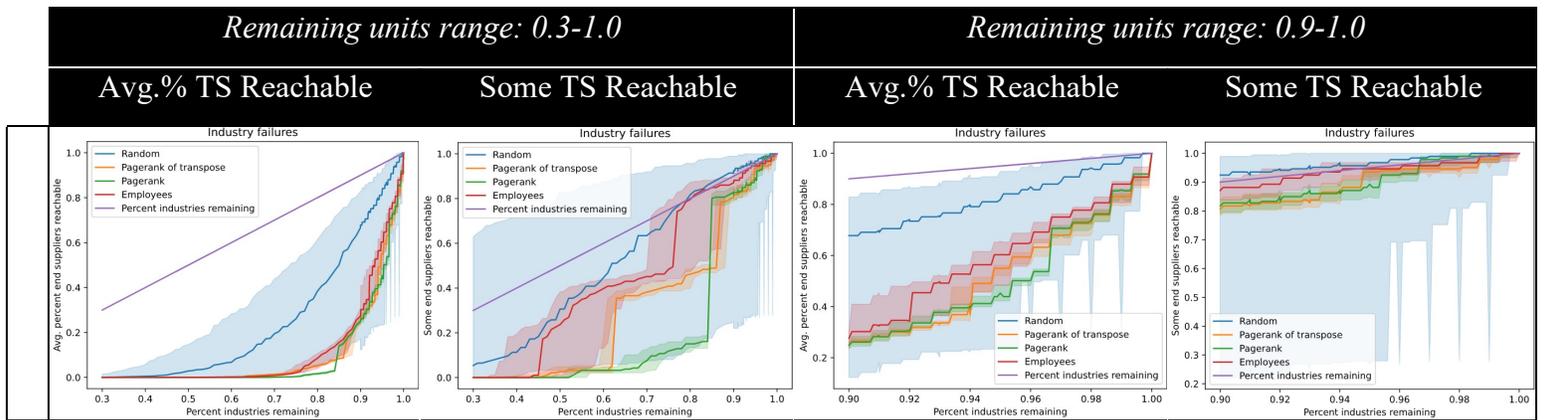

Figure 6 Caption: Industry-scale RFE & IAE of the global medical equipment supply chain.
Figure 6 Alt Text: A four-panel image of RFE & IAE at the industry-scale displays disruption's effect on the supply chain.

### 5.5. Country-scale Supply Chain: RFE and IAE Analysis

Our analysis across different scales reveals that as we move from firm-scale and country-industry-scale analysis to industry-scale and country-scale analysis, the variability of interruption resulting from random failures increases. This issue can be observed by comparing the shaded areas between the 97.5 and 2.5 percentile of outcomes. According to this, the firm-scale analysis result has the lowest uncertainty. Figure 7 displays the result of the analyses conducted at the country scale.

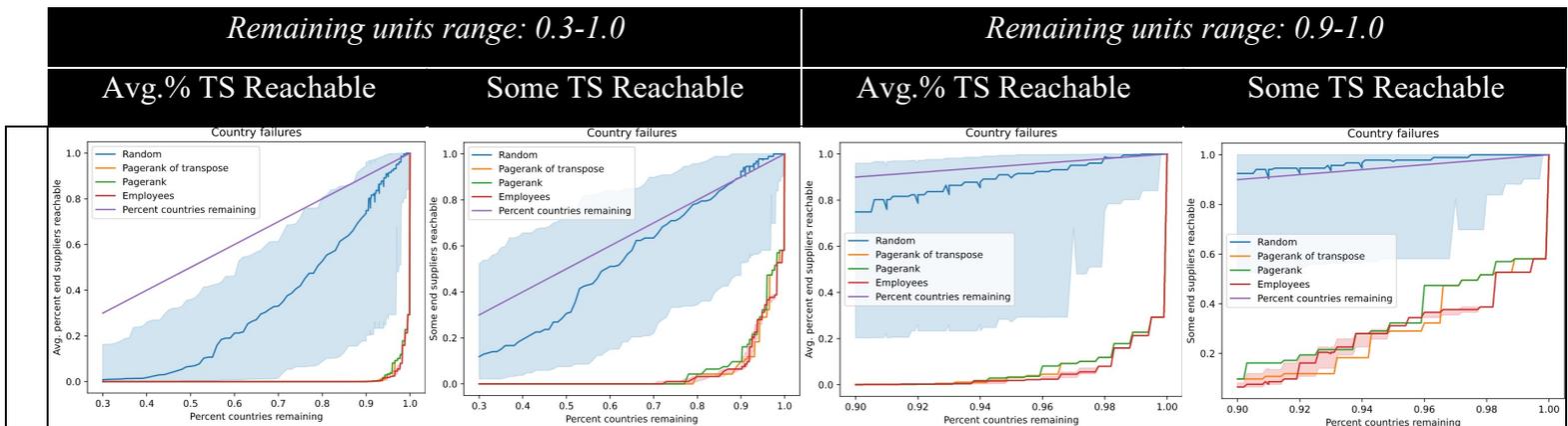

Figure 7 Caption: Country-scale RFE & IAE of the global medical equipment supply chain.
Figure 7 Alt Text: A four-panel image of RFE & IAE at the country-scale displays disruption's effect on the supply chain.

Our analysis shows the important role of some major countries like the U.S., China, and a few other countries whose elimination can create significant disruption to the global supply chain. This issue can be observed in the IAEs in Figure 7. When the first major country is eliminated based on any of the IAEs, we can observe that the reachability of intelligent attacks starts to drop to approximately 25% for ATSR and 60% for STSR.

*5.6. Supply Chain Breakdown and Fragmentation.*

Another interesting finding of this study regards the threshold of global supply chain breakdown or fragmentation at different scales. To the best of our knowledge and as reported in the literature (c.f. Wang et al. 2018), there is no inherent threshold number of firms remaining in operation that will result in the complete breakdown of global supply chain operations. Thus, we consider instead different degrees of breakdown in terms of the ATSR achieved by various attacks, comparing this to classical, percolation-theoretical estimates of breakdown.

- **Supply Chain Breakdown**

Based on our global medical equipment models, effective breakdown thresholds can be estimated from RFE and IAE results (Figures 4, 5, 7, and 7) using a user-selected minimum ATSR criterion. Table 3 summarizes the thresholds based on the supply chain breakdown where the "breakdown threshold" is defined as the largest value of percent firms remaining at which ATSR was less than 20% or 1%. Depending on the researchers' needs, other limits may be chosen (Lee et al. 2019, Rapisardi et al. 2018).

Table 3. Supply chain breakdown threshold of global medical equipment.

| Failure/Attack Type | Scale | Supply Chain Breakdown Threshold | | | |
|---|---|---|---|---|---|
| | | 20% limit | | 1% limit | |
| | | remaining | affected | remaining | affected |
| Random | Firm | 0.73 | 0.27 | 0.42 | 0.58 |
| | Country-industry | 0.72 | 0.28 | 0.41 | 0.59 |
| | Industry | 0.71 | 0.29 | 0.38 | 0.62 |
| | Country | 0.60 | 0.40 | 0.29 | 0.71 |
| PageRank of transpose | Firm | **0.93** | **0.08** | **0.84** | **0.16** |
| | **Country-industry** | **0.98** | **0.02** | **0.91** | **0.09** |
| | **Industry** | **0.88** | **0.12** | **0.66** | **0.34** |
| | **Country** | **0.99** | **0.01** | **0.94** | **0.06** |
| **PageRank** | Firm | 0.97 | 0.03 | 0.91 | 0.10 |
| | Country-industry | 0.99 | 0.01 | 0.90 | 0.10 |
| | Industry | 0.88 | 0.12 | 0.76 | 0.24 |
| | Country | 0.99 | 0.01 | 0.94 | 0.06 |
| **Employees** | **Firm** | **0.89** | **0.11** | **0.77** | **0.23** |
| | **Country-industry** | **0.98** | **0.02** | **0.92** | **0.08** |
| | **Industry** | **0.87** | **0.13** | **0.70** | **0.30** |
| | **Country** | **0.99** | **0.01** | **0.94** | **0.06** |

As illustrated in Table 3, the supply chain breakdown thresholds are dependent on the type of disruption, scale of analysis, and desired breakdown limit values (here, calculated based on 20% limit and 1% limit). For example, when only 27% of the firms are affected by a random failure, the ATSR of the global supply chain for MSFs falls down to 20%. If the breakdown limit is defined at the 1% limit, the breakdown can be achieved when 58% of firms are randomly affected.

Overall, the intelligent attacks are found to be notably more efficient in achieving supply chain breakdown limits. For example, a country-industry PageRank-based attack can achieve the supply chain breakdown limit of 20% by targeting merely 1.3% of country-industries, while achieving the same level of supply damage through random failure calls for 28% of country-industries to be eliminated from the global supply chain. The comparison of PageRank- and PageRank-of-transpose-based attacks also provides interesting results. The attacks at industry-scale and country-scale have the same level of efficiency in both types of attacks. However, PageRank is found to be marginally more effective at the firm and country-industry-scale. While PageRank provides more weight on the importance of lower tier suppliers, from the perspective of PageRank of transpose, higher tier suppliers are viewed to be more important. According to this result, eliminating lower tier suppliers is more efficient in achieving supply chain breakdown limits than eliminating higher-tier suppliers. While this result applied to the sample of medical equipment global supply chains, in other industries the supply chain breakdown thresholds may exhibit different patterns.

- **Supply Chain Fragmentation**

In addition to the abovementioned method, we wish to compare to a fragmentation threshold identification methodology from the field of graph theory based on the Erdős–Rényi (ER) model (Erdős and Rényi 1959) and power-law graphs. The reachability statistics used in this paper are tailored to the case of supply chains, but they do bear a superficial similarity to more classical "network robustness" analyses that have roots in percolation theory (c.f. Bunde and Havlin 1996). In such analyses, undirected graphs are generally assumed to have been drawn from a random graph model, and the goal is to determine, in the limit of infinitely large graphs, what proportion of nodes must be removed either randomly or in a targeted manner, in order to break up the graph into many connected components. Classical percolation theory does not immediately apply to directed graphs, nor is our preceding analysis meant to imply anything about whether the supply chain consists of multiple connected components. To visualize the difference, we imagine the supply chain laid out in tiers, with each tier occupying a single layer (See Figure 8).

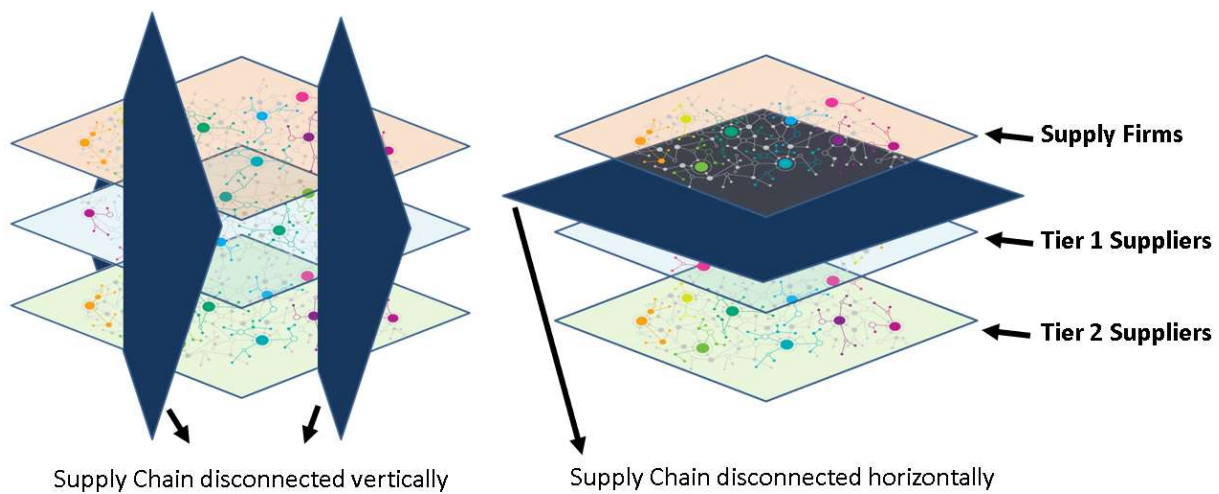

Figure 8 Caption: Supply Chain network segmentation across tiers.
Figure 8 Alt Text: Two images of multi-tier networks. One multi-tier network is sliced vertically. The other multi-tier network is sliced horizontally.

If we cut (disconnect) the supply chain horizontally, all our reachability statistics are zero, but the network only contains two giant components. Conversely, if the chain is sliced vertically several times, our reachability statistics can be quite high, even though the supply chain is broken into many pieces. Indeed, the vertical slicing could correspond to competing business ecosystems which supply among themselves but not with each other, which is a feature of modern supply chains. In such a case, the reachability statistics would be unaffected, but the network would be highly fragmented from a percolation-theoretical perspective. Despite the differences between reachability and component-based approaches, it is interesting to compare our reachability-based results with what might be obtained using more classical percolation tools. We limit ourselves to the four most popular analyses: Erdos-Renyi vs. power law graphs and random vs. (degree) targeted attacks. Note that random and targeted attacks are approximately equally effective for ER graphs.

In graph theory, network robustness is measured by assessing "the impact of node failure on the integrity of a network" (Barabási, 1999). This method is widely utilized in statistical physics and mathematics within the context of percolation theory (c.f. Bunde and Havlin 1996; Zheng et al. 2021). Based on percolation theory principles we explore the change in the structure of the network when nodes or edges are removed from the network. We start removing the supply chain nodes until the average degree of each node is less than 1 (corresponding to the ER robustness limit). We use the ER's connectedness threshold to identify the critical threshold of network fragmentation. In this study, we refer to this network fragmentation as *supply chain fragmentation*. The result of this analysis is presented in Table 4.

Table 4. Supply chain fragmentation threshold of global medical equipment.

|  | Scale | Supply Chain Fragmentation Threshold | |
|---|---|---|---|
|  |  | Avg. node degree <1 | |
|  |  | Remaining | Affected |
| Random failure | Firm | 0.21 | 0.79 |
|  | Country-industry | 0.19 | 0.81 |
|  | Industry | 0.02 | 0.98 |
|  | Country | 0.01 | 0.99 |

We define the supply chain fragmentation as the situation where the supply chain network is broken into many disconnected components, identified by the average degree falling below 1. Albert et al. (2000), in their work "Error and attack tolerance of complex networks" explain that for ER networks, targeted and random failures are about equally effective at fragmenting the network. The result from Table 4 shows that the robustness of the supply chain network measured using the ER model varies across different scales. The primary reason for this behavior is that, as we move our scale from firm and country-industry to industry and country, the network density is so great that it is hard to get the degree below one. The supply chain of medical equipment includes a relatively small number of extremely well-connected nodes and hence exhibits a highly skewed degree distribution. This characteristic makes this supply chain particularly robust to random disruptions, as the probability of failure in small well-connected firms is relatively lower. However, the same characteristic makes the supply chain highly vulnerable to intelligent attacks due to the high impact of targeted attacks launched on the small number of the well-connected firms.

If instead we consider the supply chain as a power law network, we use the Molly-Reed criterion and an estimate of 1.4 for the power law exponent (obtained using the powerlaw python package). Thus, for random attacks all but .8% of firms must be deleted in order to break up the network. In the targeted case, the theory predicts that arbitrarily small attacks should break up the network. This is broadly consistent with the extreme observed efficacy of targeted attacks on the real network, which is expected, since the actual degree distribution is quite heavy tailed. While there is a consensus in the literature that "supply chains are often highly vulnerable in general" however the extent of vulnerability to random failures and intelligent attacks was not previously measured using real work data (Trkman and McCormack, 2009; Tang et al., 2016). This work provides notable advancement toward the application of new methodologies in measuring the interruption thresholds.

## 6. Discussion

Ivanov and Dolgui (2020), as some of the seminal scholars in the field, promoted the study of viability analysis based on the intertwined network of supply chains. Ivanov and Dolgui (2020) highlighted the different behaviors of intertwined supply chain networks versus the traditional linear supply chains. As we identified in the complex network of about 150,000 supply chain connections, there exist numerous supply chain loops. As we expanded mining supply chain data tier by tier, we also identified numerous lower tier suppliers that become suppliers to the higher tier firms. These firms create intertwined value co-creation business ecosystems forming what Dolgui et al. (2020) refer to as "value webs".

Overall, the global supply chain network of medical equipment exhibits high vulnerability exhibited by a sharp decrease in TSs reachable to MSFs when the supply chain is faced with disruptions. The high scale of vulnerability is due to the very few alternative routes from higher-tier suppliers to the MSFs. The reason for such a vulnerable network structure is twofold. On the one hand, the medical equipment supply chain requires incorporating a high scale of service (Maltz and Maltz, 1998) due to its final products' complexity and sensitive nature. This factor limits the flexibility of MSFs to maintain costly supplier relationships with multiple suppliers simultaneously. On the other hand, over the past few decades, efficiency goals (NASEM, 2018; Jha, 2019) have "forced" (Denton and Jaska, 2014) MSFs to adopt creative efficiency practices and strategies including "pull," "push," "just-in-time (JIT)," "economies of scale" and "off-shoring."

The result of this study reveals that disruption in a small number of suppliers across any of the analyzed tiers can have a devastating effect on the supply of medical equipment. In such a business environment, one of the key questions facing supply chain managers is how many tiers of the supply chain need to be analyzed to obtain enough information about the structure of the supply chain? We provided a data-driven method that identified the minimum number of tiers to be analyzed to acquire such information. We provided arguments and evidences (c.f. Davis, 2021[a] and 2021[b]) that firms should monitor their supply chain beyond the first tier. We recommend practitioners to use the proposed convergence tool and to identify and include the convergence tier in their monitoring procedure.

The analysis of the effects of disruptions in this paper is primarily based on ATSR and STSR. However, it is noteworthy to mention that we also computed the percent of MSFs with All Terminal Suppliers Reachable (ALTSR). Through our experiments, we observed that ALTSR drops to zero with high probability almost immediately because each MSF depends on thousands of firms. This exhibits one of the limitations of the study, as in real-world operations, firms usually carry buffer inventories and hence may be able to identify alternative suppliers across different tiers. Nevertheless, such extreme experiments can provide beneficial information to identify critical supply paths, considering the high dependencies in the medical equipment's global supply chain. In addition to supply chain tier analysis and random vs. intelligent disruptions, we also provided a novel approach to measure and illustrate the thresholds of supply chain breakdown and fragmentation.

We also tried to answer the question of when a firm's supply chain breaks down and collapses. While the breakdown threshold varies from industry to industry and from firm to firm, we proposed a methodology that can be used by each firm or each industry based on their self-determined limit. To illustrate our methodology, we simulated supply chain breakdown effects at 1% and 20% limits (Table 3) and supply chain fragmentation at ER limit (Table 4) using our real-world data. In general, the supply chain network of medical equipment exhibited relatively more robustness to random failures while it was notably fragile against intelligent attacks.

As we move from the firm scale network (with the largest number of nodes) to the county scale network (with the smallest number of nodes), two opposite forces affect the robustness of the supply chain. On the one hand, in larger networks, the probability of eliminating a more central node at each marginal level of disruption is lower. Therefore, we would expect (under the ceteris paribus condition) the supply chain threshold to be lower for smaller networks, *i.e.*, industry scale and country scale networks. This argument is valid on the assumption that in larger networks (e.g., firm-level), a lower percentage of firms is highly central, which is true in our sample of medical equipment supply chain networks.

On the other hand, to produce more macro-level networks, we aggregate the edges from the previous scale. Consequently, as we move from firm scale disruptions to country scale disruptions, the models with a smaller number of nodes assumes a higher level of supply chain (node) substitution in the network. For example, the model may assume an industry or a country can substitute the elimination of a node, which does not necessarily apply to all cases of interruptions in the real world. These two opposite forces can explain variances in the breakdown thresholds. To illustrate the threshold of supply chain fragmentation, we used the commonly used threshold in the network science literature corresponding to the ER robustness limit (Table 4). Similar to supply chain breakdown analysis, the two opposite forces are at play to explain the results. The result indicates greater robustness of the network at industry and country scales.

The global supply chain of medical equipment is comprised of interdependent networks of suppliers from different industries and countries. Such "interdependent networks are difficult to defend by strategies such as protecting" the highly central firms or industries (Huang et al. 2010). One solution that increases resiliency to attacks and failures and minimizes the cascading effect of failures is to focus on protecting clusters/communities of firms (business ecosystems) instead of only protecting the central firms in order.

Another contribution of this work is its application in the validation of future simulation algorithms. Our analysis provides benchmark supply chain patterns of behavior to be used in future simulation algorithms to produce more accurate "in silico" models that can better "mimic real data" (De Smet and Marchal, 2010). The findings and methodologies utilized in this study have notable implications for policy makers working on commerce, national security and public health. Also, relying on multiple MSFs does not necessarily diversify the risk as many MSFs share some supply pathways across multiple tiers of backward supply chain. These are some of the topics that policy makers can strategically explore using such network analytics models.

7. **Implications for research and practice**

As reviewed by Perera et al. (2017), various attempts have been made to apply network science ideas to understanding the supply network and its robustness. The majority of these attempts have focused on describing the network topology, especially noting the heavy-tailed degree distribution. Various models have been proposed to generate artificial supply chain networks, including Barabasi-Albert models and fitness-based models. Notably, most of these models only produce undirected networks. Network robustness metrics have then been applied to these synthetic networks, largely drawing inspiration from computer networking theory. Generally, the network is disrupted and one of the robustness metrics, such as the size of the largest connected component after disruption or average path length in the largest connected component, is tracked as a function of the number of remaining nodes. Using ideas from percolation theory, it is possible to obtain analytical results for some graph models. The general theme of such analyses is that random attacks are not very disruptive, but even small degree-targeted attacks are devastating.

The network science approaches have contributed insight by noting the importance of the undirected topology for understanding robustness, but a number of key gaps remain unfilled by this literature. First, most supply chain network models are still too unrealistic, generally not even modeling edge direction. Second, the network robustness metrics are copied from computer science and don't necessarily encode the actual robustness of supply chains, which don't need a (weakly) connected component or short paths to function effectively. Our work addresses these gaps by (1) using the most complete, real supply chain network available and (2) using a more realistic assessment of supply chain functionality in terms of suppliers' ability to reach their customers. While the percolation-theoretic arguments may no longer directly apply with these modifications, the added realism is a significant advantage to our formulation.

The present study has notable contributions for practitioners as well as policymakers. There have been several calls for enhancing the organizational supply chain visibility beyond the first-tier suppliers. This issue is further underscored by recognizing that contemporary supply chains are made up of intertwined networks that expand globally across different industries. From a supply chain strategy perspective, businesses need to have a realistic map of their supply chain to show their dependencies on other firms, industries, and countries throughout multiple tiers.

One question facing managers is: how many tiers of supply chain firms should firms monitor to ensure they have proper analysis depth? We provided an analytical tool, which enables firms to know the required number of tiers they need to explore (convergence tier) to ensure they have the required visibility to their supply chain. Once the supply chain network data is collected up to the convergence tier, businesses can reliably use the data to conduct clustering analysis to identify the boundaries of their business ecosystem and their competitors' business ecosystems. Another important application of this study for businesses is that firms can identify their vulnerability to random failures and intelligent attacks at four different scales, as discussed before. Our model is a ready-to-use tool for firms to identify the cascading consequences of such interruptions in a timely manner.

Furthermore, in the cases of such interruption, our large-scale real-world network data can show the alternate possible network paths. Since our data only includes notable suppliers, as explained in the data collection section, the solutions will only include such firms. Our analyses showed that the global supply chain of medical equipment could be highly influenced by a small number of highly connected firms. Firms operating in this industry are recommended to identify these highly central firms and invest in the sustainability of their supply chain connections with these central firms. One of the interesting applications of this study is in asset management. While some of the applications of this study were discussed for the firms operating in the industry (e.g., manufacturers and suppliers), our result provides notable information about the vulnerability of firms, industries, and countries to different types of disruptions. A political unrest in a country, a natural disaster affecting a particular industry in a specific country, a global/nationwide raw material shortage that affects an industry, or the bankruptcy of a major company can affect the asset value of firms across different supply chain tiers. Our vulnerability assessment methodology allows investors and asset managers to measure the risk of interruptions and hedge their investments. In particular, our breakdown and fragmentation analysis methodologies enable asset managers to measure their risk levels using their defined threshold limits at any of the four scales of analysis.

The present work also has critical application for policymakers. The COVID-19 pandemic illustrated how much the national security and well-being of every country's citizens can be negatively affected by such supply chain disruption. More importantly, citizens and governments became aware of the high level of fragility of global supply chains. One question facing policymakers is what are the most central firms, industries, and countries in medical equipment supply? Our analysis provides the list of most critical firms, industries, and countries across the ten tiers of the medical equipment supply chain. The same data collection and analysis can be applied to other sectors to identify strategically essential firms and industries to each country's national security. This methodology can be applied as an attack and a defense plan for firms as well as the government. Our analysis identified the catastrophic effect of intelligent attacks on the supply of medical equipment. Competitors (i.e. competing firms) and political adversaries (e.g. terrorists and cyber criminals) could target the highly central firms or industries to cause the most damage. As a countermeasure, firms and policymakers can use our findings to identify the highly central firms. Protecting the relatively small number of highly connected firms can ensure the firms' competitiveness and secure the supply of medical equipment to citizens.

The findings as they relate to the mapping of the supply chain can be utilized in identification and tracking of labor exploitation (c.f. LeBaron, 2021) and modern slavery (Sokat and Altay, 2022). The supply chain mapping further provides opportunities to not only explore business continuity planning (c.f. Zsidisin et al., 2003), sustainability (c.f. Mubarik et al. 2021) and cybersecurity (c.f. Melnyk et al. 2022), but also plan for preventative and contingency plans. While the supply chain mapping can have wide applications in different domains, we believe our proposed RFE and IAE in particular has notable application in cybersecurity and protecting the national supply chains (c.f. Mubarik et al. 2022).

Finally, this study demonstrated the feasibility of mapping the supply chain using mined financial information. While this method of data collection has some limitations, it is arguably the most feasible and comprehensive way to provide a "macro view of the supply chain structure" (c.f. Ivanov and Dolgui, 2021) at the global level. Additionally, this method is affordable and less laborious, which can be of great importance to SMEs with limited resources. For example, once the coding is completed, tested, and debugged (which can take a few months for the first batch of data) moving forward, an operator can mine and analyze a multi-tier supply chain for any industry in a few days. However, this process can be automated with cloud high-performance computing (HPC) to produce results in a shorter time.

## 8. Limitations and Directions for Future Studies

Some of the constraints of this study are related to our data. As previously discussed the network is limited to notable suppliers and the monetary value of transactions is not available. 99% of firms in the OCED countries are SMEs (Lin et al. 2022). While our study includes data from all countries, a representative sample ideally—and arguably—is expected to include closer to 99% SMEs. Due to the data mining methodology, a privately owned firm will not be included in our sample if the firm does not have notable supply chain transactions with a qualifying firm (that is, a firm that is required to file financial reports with the SEC). 44% of economic activity in the U.S. is generated by small businesses, and they play a critical role in the nations' economic prosperity (SBA, 2019). As discussed in the data collection section, at least 52% of firms in our sample are SEMs. If we assume that firms with missing employee data are privately owned SEMs (which is a reasonable assumption), the percentage of SMEs in our sample can be estimated to be up to 69%. In general, SMEs and private firms have lower representation in our sample vis-à-vis the representation of these firms in the global population of firms active in the supply chain of medical equipment. Underrepresentation of SMEs is one of the limitations of the study, particularly since SMEs are generally considered to be the supply chain's weakest links (CERT-UK, 2015). There is evidence that not only do smaller businesses face more volatility stemming from disruptions such as COVID-19 (Giunipero et al., 2022), but also they are less prepared and equipped (operationally and financially) to navigate and survive such vulnerabilities (OECD study, 2009).

Other limitations of the present study stem from our operational assumptions, as discussed previously (Table 1). One of the operational assumptions is that companies can only substitute their suppliers with the existing firms in the network through the existing paths, and furthermore, any alternative path can be used as an alternative supply path for any firm. The fact is that "replacing suppliers is not always feasible," especially for hard to replace items (Duman et al. 2022) and supplies with a low degree of substitutability. Replacing suppliers can be costly and time-consuming due to several factors, including operating constraints (Duman et al., 2022), the requirements for "product redesign" (Hansen and Schmitt, 2021), and geographic location (Kleifgen et al., 2022).

Other operational limitations are associated with the assumed absence of capacity limit for each supply path, the absence of time lag, and the supply chain being limited only to notable suppliers and customers. We recognize that these assumptions are not realistic in many cases. Future studies can further study this subject by removing any or a number of these operational assumptions, pending the availability of data.

Since we tried to use the most efficient algorithms, we did not face significant computational challenges that may arise from the data's size and complexity. While we could analyze the data across ten tiers for this industry using high-performance personal computers, we recommend researchers to consider using an affordable, commercially available cloud HPC for analyzing larger networks.

The present study explored the backward supply chain. Future studies in the area are encouraged to conduct forward supply chains. The study of supply chains across multiple tiers and scales can be conducted at the network level, the process level and the control level (Ivanov and Dolgui, 2021). This study has been on the network-level, and there are fruitful opportunities for expanding the present work to process-level and control-level.

While the current research focuses on analyzing nodes, we believe there are fruitful research opportunities in conducting clustering analysis based on community detection algorithms. Future studies are encouraged to explore supply chains' vulnerability when a cluster of firms is affected or when central firms across different communities are affected. Furthermore, this study is based on the analysis of the supply chain networks. Future studies can explore the global supply chains using multi-layer connected networks of business ecosystems.

In this study, we proposed analytical tools to identify the survival of firms based on access to suppliers. In future studies, we plan to identify the firms that survived some level of failure or attacks. We are interested in exploring various characteristics of firms to identify potential supply chain, firm specific and/or industry-specific characteristics that can contribute to higher probability of enduring such events. One such domain can be the study of relationship between uncertainty and profitability (c.f. Knight, 1921). While most studies in this domain explore the effect of disruption, there has been a gap in the literature on exploring post-disruption supply chain management (Ivanov, 2021). The research methodologies employed in this paper are compatible with additions to simulate random as well as intelligent recovery based on "percolation strategies" (c.f. Smith et al. 2019). We plan to conduct such studies in the future with the goal of identifying the optimized recovery paths.

**Disclosure statement**

No potential conflict of interest was reported by the authors.

**Acknowledgments**

- We would like to acknowledge and appreciate the valuable and constructive feedback provided by the anonymous reviewers.
- This project is part of the NCCU Advanced Center for COVID Related Disparities (ACCORD) supported by the North Carolina Policy Collaboratory at the University of North Carolina at Chapel Hill with funding from the North Carolina Coronavirus Relief Funds and University of North Carolina at Chapel Hill (US).
- We acknowledge Peter Mucha, David Grimsman, Sean Warnick, Tyler Burrows, Benjamin Webb and Mike Aguilar for helpful discussions.
- The first and second authors contributed equally.

**Supplementary files on GitHub**

- Python codes included as supplementary material.
- Dataset of the top 100 firms, industry-countries, industries, and countries with the highest in-degree, out-degree and PageRank and impact on supply chain reachability.

**GitHub**: https://github.com/businessecosystem

**Data Availability Statement**

The raw data was mined from proprietary databases and prepared by IEBE Lab housed at North Carolina Central University. Derived data supporting the findings of this study that includes notable results of the supply chain disruption analysis at each of the four scales of analysis is available on GitHub. Further derived data supporting the findings of this study can be made available from the corresponding author (K. L.) on request.

**References**

Abdulla, B. and B. Birgisson 2020. "Characterization of Vulnerability of Road Networks to Random and Nonrandom Disruptions Using Network Percolation Approach." Journal of Computing in Civil Engineering, 35(1): 1-13. https://doi.org/10.1061/(ASCE)CP.1943-5487.0000938

Abele E., J. Elzenheimer, T. Liebeck and T. Meyer. 2006. "Globalization and Decentralization of Manufacturing." In: Dashchenko A.I. (eds) *Reconfigurable Manufacturing Systems and Transformable Factories*. Springer, Berlin, Heidelberg.

Albert, R., H. Jeong, and A. L. Barabási. 2000. "Error and attack tolerance of complex networks." *Nature* 406: 378–382 (2000). https://doi.org/10.1038/35019019

Aleksiejuk, A., and J. A. Holyst. 2001. "A simple model of bank bankruptcies, Physica A 299: 198–204.

Ash, J. and D. Newth. 2007. "Optimizing complex networks for resilience against cascading failure." *Physica A: Statistical Mechanics and its Applications* 380(1): 673-683.

Barabási, A.L. 1999. "Network Science." http://networksciencebook.com/, Last Accessed 4/10/2021.

Barabási, A-L., Albert, R. 1999. "Emergence of scaling in random networks. Science. 286 (5439), pp: 509–512.

Barratt M. and, A. Oke. 2007. "Antecedents of supply chain visibility in retail supply chains: A resource-based theory perspective." *Journal of Operations Management* 25 (6):1217 – 1233.

Barratt, M. and R. Barratt. 2011. "Exploring internal and external supply chain linkages: Evidence from the field." *Journal of Operations Management* 29 (5): 514 – 528.

Barrow H.J. 2019. "NETWORK SHAPING." *Master of Science in Operations Research*, Naval Postgraduate School. URL https://apps.dtic.mil/dtic/tr/fulltext/u2/1080001.pdf [Last accessed 11/05/2020]

Baz E. E., and S. Ruel. 2021. "Can supply chain risk management practices mitigate the disruption impacts on supply chains' resilience and robustness? Evidence from an empirical survey in a COVID-19 outbreak era." *International Journal of Production Economics* 223: 107972.

Bhamra, R., S. Dani, and K. Burnard. 2011. "Resilience the concept, a literature review and future directions." *International Journal of Production Research*, 49(18): 5375-5393.

Blackhurst, J. and T. Wu. 2009. "Book Introduction." In (Eds.) T. Wu and J. Blackhurst, *Managing supply chain risk and vulnerability: Tools and methods for supply chain decision makers*, Springer, London.

Bommel, W.V. 2010. "A conceptual framework for analyzing sustainability strategies in industrial supply networks from an innovation perspective." *Journal of Cleaner Production,* 19(8): 895-904.

Boston, W. 2022. "Ukraine War Plunges Auto Makers Into New Supply-Chain Crisis." *Wall Street Journal*, 2022/03/03. URL: https://www.wsj.com/articles/ukraine-war-plunges-auto-makers-into-new-supply-chain-crisis-11646309152 [Last accessed, 21/08/2022]

Broadbent, S. and J. Hammersley. 1957. "Percolation processes I. Crystals and mazes." *Mathematical Proceedings of the Cambridge Philosophical Society* 53 (3): 629–641.

Brown, G. G., W. M. Carlyle, J. Salmeron and K. Wood. 2005. "Analyzing the Vulnerability of Critical Infrastructure to Attack and Planning Defenses." *Tutorials in Operations Research*: 102-123.

Bunde, A. and S. Havlin. (1996) *Fractals and Disordered Systems* (Springer, New York.

Butt, A. S. 2021. "Strategies to mitigate the impact of COVID-19 on supply chain disruptions: a multiple case analysis of buyers and distributors." *The International Journal of Logistics Management,* xx(xx), xx-xx-. https://doi.org/10.1108/IJLM-11-2020-0455

Cassens, I., P. Mardulyn, M. C. Milinkovitch 2005. "Evaluating Intraspecific "Network" Construction Methods Using Simulated Sequence Data: Do Existing Algorithms Outperform the Global Maximum Parsimony Approach?" *Systematic Biology* 54(3): 363-372.

Chan, H. K. and F. T. S. Chan. 2010. "Comparative study of adaptability and flexibility in distributed manufacturing." *Decision Support Systems* 48(2): 331-341.

Choi T. and T. Linton. 2011. "Don't let your supply chain control your business." *Harvard Business Review*, 89 (12): 112–117

Cliquet, G. and, P. A. Guillo. 2013. "Retail network spatial expansion: An application of the percolation theory to hard discounters." *Journal of Retailing and Consumer Services,* 20(2): 173-181.

Crosignani, M., Macchiavelli, M. and Silva A. F. 2021. "Cyberattacks and Supply Chain Disruptions." *Liberty Street Economics,* 2021/06.22, Federal Reserve Bank of New York. URL: https://libertystreeteconomics.newyorkfed.org/2021/06/cyberattacks-and-supply-chain-disruptions/ [Last accessed, 08/20/2022].

Cui, H., Loureiro, B., Krzakala, F., Zdeborova, L. 2022. "Error Rates for Kernel Classification under Source and Capacity Conditions." Machine Learning, arXiv:2201.12655

Davis, R. (2021[a]). How Toyota Dodged the Chip Shortage. "*Business Week*." April 12, 2021. URL: https://www.magzter.com/stories/Business/Bloomberg-Businessweek/How-Toyota-Dodged-The-Chip-Shortage [Last accessed, 3/15/2022]

Davis, R. (2021[b]). How Toyota Steered Clear of the Chip Shortage Mess. "*Business Week*." April 7, 2021. URL: https://www.bloomberg.com/news/articles/2021-04-07/how-toyota-s-supply-chain-helped-it-weather-the-chip-shortage [Last accessed, 3/15/2022]

De Smet, R. and K. Marchal. 2010. "Advantages and limitations of current network inference methods." *Nature Reviews Microbiology,* 8: 717–729.

Delic, M. and D. Eyers. 2020. "The effect of additive manufacturing adoption on supply chain flexibility, and performance: An empirical analysis from the automotive industry." *International Journal of Production Economics*, 228: 107689.

Denton, T. and P. Jaska. 2014. "Managing the medical device segment of the healthcare supply chain." *International Journal of Business and Public Administration* 11(2): 41-47.

Dolgui, A., D. Ivanov and B. Sokolov. 2020. Reconfigurable supply chain: The X-Network. *International Journal of Production Research*. 58(13): 4138-4163.

Duman, O., L. Wang, M. Au, M. Kassouf and M. Debbabi M. 2022. "Hardening Substations against Supply Chain Attacks Under Operational Constraints," 2022 IEEE Power & Energy Society Innovative Smart Grid Technologies Conference (ISGT), 2022, pp. 1-5, doi: 10.1109/ISGT50606.2022.9817488.

Erdős, P. and A. Rényi. 1959. "On random graphs I." *Publ. Math. Debrecen* 6: 290–297.

Fierro, L. H., R. E. Cano, and J. I. Garcia. 2020. "Modelling of a multi-agent supply chain management system using Colored Petri Nets." *Procedia Manufacturing* 42: 288-295.

Ghadge, A., Er Merve, D. Ivanov and A. Chaudhari. 2021. "Visualisation of ripple effect in supply chains under long-term, simultaneous disruptions: a system dynamics approach", *International Journal of Production Research*, https://doi.org/10.1080/00207543.2021.1987547.

Grida, M., R. Mohamed, and A. N. H. Zaied. 2020. "Evaluate the impact of COVID-19 prevention policies on supply chain aspects under uncertainty." *Transportation Research Interdisciplinary Perspectives* 8: 1-8.

Giunipero, L.C., D. Denslow, A.I. Rynarzewska. 2021. Small business survival and COVID-19 - An exploratory analysis of carriers." *Research in Transportation Economics*, 93, https://doi.org/10.1016/j.retrec.2021.101087

Guo, H., S. S. Yu, H. H. C. Iu, T. Fernando and C. Zheng. 2019. "A complex network theory analytical approach to power system cascading failure—From a cyber-physical perspective." *Chaos: An Interdisciplinary Journal of Nonlinear Science*, 29(5): 053111.

Hall, C. (2019). Companies Cite Costs for Lag in Supply Chain Mapping, Transparency. Sourcing Journal (on-line). April 1, 2019.

Hallquist, M.N., A. G. C. Wright and P. C. Molenaar, 2019. "Problems with Centrality Measures in Psychopathology Symptom Networks: Why Network Psychometrics Cannot Escape Psychometric Theory." *Multivariate Behavioral Research* 56(2): 199-223.

Hansen, E.G. and J. C. Schmitt. 2021. "Orchestrating cradle-to-cradle innovation across the value chain." *Journal of Industrial Ecology* 25(3), pp. 627-647. DOI: 10.1111/jiec.13081

Haque, H., S. K. Paul, R. Sarker and D. Essam. 2020. "Managing decentralized supply chain using bilevel with Nash game approach." *Journal of Cleaner Production*, 266(1): 121865.

Hausman, A.V. 1998. "Cooperative adoption of technological innovation in the context of long-term relationships: A channel perspective." PhD Dissertation, May 1998, University of South Florida. https://www.proquest.com/docview/304454782 [Last accessed 9/20/2021]

Helbing D. and C. Kuhnert. 2003. "Assessing interaction networks with applications to catastrophe dynamics and disaster management." *Physica A* 328: 584–606.

Hernandez, J.M., and C. Pedroza-Gutierrez. 2019. "Estimating the influence of the network topology on the agility of food supply chains." *PLoS One*, 14(7): 1-21.

Hill, R., Sarkani, S. and T. A. Mazzuchi. 2021. "Managing in a Post-COVID-19 World: A Stakeholder Network Perspective, IEEE Engineering Management Review, 49(1): 63-71.

Hong, S., B. Wang, X. Ma, J. Wang, and T. Zhao. 2015. "Failure cascade in interdependent network with traffic loads." *Journal of Physics A: Mathematical and Theoretical*, 48, pp: 1-12.

Hosseini, S.M. and Ivanov D. 2021. "A multi-layer Bayesian network method for supply chain disruption modelling in the wake of the COVID-19 pandemic." *International Journal of Production Research*, DOI: 10.1080/00207543.2021.1953180

Ismail, H.S. and H. Sharifi. 2006. "A balanced approach to building agile supply chains." *International Journal of Physical Distribution & Logistics Management* 36(6): 431-444.
Ivanov, D. 2021. Exiting the COVID-19 pandemic: after-shock risks and avoidance of disruption tails in supply chains, *Annals of Operations Research*, https://doi.org/10.1007/s10479-021-04047-7.
Ivanov, D., and A. Dolgui. 2020. "Viability of Intertwined Supply Networks: Extending the Supply Chain Resilience Angles towards Survivability. A Position Paper Motivated by COVID-19 Outbreak." *International Journal of Production Research*, 58(10): 2904-2915.
Ivanov, D., and A. Dolgui. 2021. "OR-methods for coping with the ripple effect in supply chains during COVID-19 pandemic: Managerial insights and research implications." *International Journal of Production Economics* 232: 107921.
Ivanov, D., B. Sokolov, A. Pavlov. 2013. "Dual problem formulation and its application to optimal redesign of an integrated production–distribution network with structure dynamics and ripple effect considerations". *International Journal of Production Research*, 51(18): 5386-5403.
Jha, S. 2019. "Trends in medical devices distribution: 2019 and beyond." *Biospectrum*, December 2019: 22-23.
Jia, F., Gong, Y. and Brown, S. (2019). Multi-tier sustainable supply chain management: The role of supply chain leadership, International Journal of Production Economics, 217: 44-63.
Joshanloo, M. 2021. "Centrality and Dimensionality of 14 Indicators of Mental Well-Being in Four Countries: Developing an Integrative Framework to Guide Theorizing and Measurement." *Social Indicators Research*. https://doi.org/10.1007/s11205-021-02723-6
Kleifgen, E., R. Duncan, and I. Stepanok. 2022. "The Covid-19 pandemic and international supply chains." IAB-Discussion Paper 202205, Institut für Arbeitsmarkt- und Berufsforschung (IAB), Nürnberg [Institute for Employment Research, Nuremberg, Germany].
Krause, S.M., M. M. Danziger, and V.Zlatic. 2016. "Hidden Connectivity in Networks with Vulnerable Classes of Nodes." *PHYSICAL REVIEW* X 6, 041022.
Krause, S.M., M. M. Danziger, and V.Zlatić. 2016. "Hidden Connectivity in Networks with Vulnerable Classes of Nodes." *Physical Review* X 6, 041022: 1-18.
Kosasih E.E. and Brintrup A. 2021. "A machine learning approach for predicting hidden links in supply chain with graph neural networks." *International Journal of Production Research*, https://doi.org/10.1080/00207543.2021.1956697
Kumar, D. and Z. Rahman 2017. "Analyzing enablers of sustainable supply chain: ISM and fuzzy AHP approach." *Journal of Modelling in Management* 12(3): 498-524.
Knight, F. H. 1921. "*Risk, uncertainty and profit.*" Boston: Houghton Mifflin.
Lavassani, K. M. 2017. "Coopetition and sustainable competitiveness in business ecosystem: a networks analysis of the global telecommunications industry." *Transnational Corporate Review* 9(4): 281-308.
LeBaron, G. (2021). "The Role of Supply Chains in the Global Business of Forced Labour." *Journal of Supply Chain Management*, https://doi.org/10.1111/jscm.12258
Lee, E, S. Emmons, R. Gibson, J. Moody, P. J. Mucha. 2019. "Concurrency and reachability in treelike temporal networks." *Phys Rev E.* 100(6-1): 062305. doi: 10.1103/PhysRevE.100.062305. PMID: 31962508; PMCID: PMC6989038

Li, W. H., Y. Han, P. F. Wang and H. Z. Guan. 2019. "Invulnerability Analysis of Traffic Network in Tourist Attraction Under Unexpected Emergency Events Based on Cascading Failure." *IEEE Access*, 7: 147383-147398.

Lin, D-Y., S. N. Rayavarapu, Tadjeddine K. and Yeon R. 2022. "Small and medium-size enterprises are facing compounding challenges. Governments and other institutions worldwide are launching programs to provide them the advisory support needed to meet the moment." *McKinsey & Company* URL: https://www.mckinsey.com/industries/public-and-social-sector/our-insights/beyond-financials-helping-small-and-medium-size-enterprises-thrive [Last accessed 8/22/2022]

Liu, J.-G., Z. T. Wang and Y. Z. Dang. 2005. "Optimization of robustness of scale-free network to random and targeted attacks." *Modern Physics Letters B* 19(16): 785–792.

Magnien, C., M. Latapy and J. L. Guillaume. 2011. "Impact of random failures and attacks on poisson and power-law random networks." *ACM Computing Surveys (CSUR)* 43(3): 1-67.

Maltz, A. and E. Maltz. 1998. "Customer service in the distributor channel empirical findings." *Journal of Business Logistics* 19(2): 103-129.

Marshall, D., L. McCarthy, P. McGrath, and F. Harrigan. 2016. "What's Your Strategy for Supply Chain Disclosure? MIT Sloan Management Review, 57(2): 37-45.

Melnyk, S. A., Schoenherr, T., Speier-Pero, C., Peters, C., Chang, J. F. and Friday D. (2022). "New challenges in supply chain management: cybersecurity across the supply chain", International Journal of Production Research, DOI: 10.1080/00207543.2021.1984606.

Melnyk, S. A., Lummus, R. R., Vokurka, R. J., Burns, L. J. and Sandor, J. (2009). "Mapping the Future of Supply Chain Management: a Delphi Study." International Journal of Production Research, 47 (16): 4629–4653.

Mena, C., Humphries, A. and Choi, T.Y. (2013). "Toward a Theory of Multi-Tier Supply Chain Management", *Journal of Supply Chain Management*, 49(2): 58- 77.

Mittendorf, B., J. Shin, and, D. H. Yoon. 2021. "Information Disclosure Policy and Its Implications: Ratcheting in Supply Chains. Journal of Marketing Research, https://doi.org/10.1177/00222437211035115 .

Mizgier, K., S. M. Wagner, and J. A. Holyst. 2012. "Modeling defaults of companies in multi-stage supply chain networks." *International Journal of Production Economics* 135(1): 14-23.

Mubarik, M.S.; Kusi-Sarpong, S.; Govindan, K.; Khan, S. A.; Oyedijo, A. (2021). "Supply chain mapping: a proposed construct", International Journal of Production Research, 17, 2021. https://doi.org/10.1080/00207543.2021.1944390

Mourtzis, D. and M. Doukas. 2013. "Decentralized Manufacturing Systems Review: Challenges and Outlook." In: Windt K. (eds) *Robust Manufacturing Control. Lecture Notes in Production Engineering*. Springer, Berlin, Heidelberg.

Nagurney, A. 2021. "Optimization of supply chain networks with inclusion of labor: Applications to COVID-19 pandemic disruptions." *International Journal of Production Economics* 235: 108080.

NASEM, (National Academies of Sciences, Engineering, and Medicine); Health and Medicine Division; Board on Health Sciences Policy." "Impact of the Global Medical Supply Chain on SNS Operations and Communications." Workshop Proceedings. Washington (D.C.): *National Academies Press* (U.S.); 2018 Jul 18.  s. Available from: https://www.ncbi.nlm.nih.gov/books/NBK525656/

Newman, M. E. 2006. Finding community structure in networks using the eigenvectors of matrices. Phys. Rev. E 74:036104. doi: 10.1103/PhysRevE.74.036104

Okeagu, C. N., D. S. Reed, L. Sun, M. M. Colontonio, A. Rezayev and Y. A. Ghaffar, R. J. Kaye, E. M. Cornett, C. J. Fox, R. D. Urman and A. D. Kaye. 2021. "Principles of supply chain management in the time of crisis." *Best Practice & Research Clinical Anesthesiology* xxx(x): xx-xx. https://doi.org/10.1016/j.bpa.2020.11.007

Peck, H. 2003. "Creating Resilient Supply Chains: A Practical Guide." Cranfield University, *Center for Logistics and Supply Chain Management*, [Last accessed 12/9/2020 URL: http://www.cranfield.ac.uk/som/scr]

Perera, S., M. G. H. Bell, and M. C. J. Bliemer. 2017. "Network science approach to modelling the topology and robustness of supply chain networks: a review and perspective, Applied Network Science, 2(33): 1-25.

Potts, M.W., P. A. Sartor, A. Johnson and S. Bullock. 2020. "A network perspective on assessing system architectures: Robustness to cascading failure." *Systems Engineering* 23(5): 597-616.

Powers, V.V. 2021. "Winter Blast to Texas Interrupts Food Supply Chain." *Ranch and Rural Living*; San Angelo, 102(6), (Mar 2021).

Pradhan S. K. and, S. Routroy. 2008. "Improving supply chain performance by Supplier Development program through enhanced visibility." *Materials Today: Proceedings* 5: 3629–3638

Rampfl, S. 2013. Seminars FI / IITM / ACN SS2013, *Network Architectures and Services*, August 2013.

Rapisardi, G., G. Caldarelli and G. Cimini. 2019. Numerical Assessment of the Percolation Threshold Using Complement Networks. In: Aiello L., Cherifi C., Cherifi H., Lambiotte R., Lió P., Rocha L. (eds) *Complex Networks and Their Applications VII.* COMPLEX NETWORKS 2018. Studies in Computational Intelligence, vol. 812. Springer, Cham. https://doi.org/10.1007/978-3-030-05411-3_65

Remko, v.H. 2020. "Research opportunities for a more resilient post-COVID-19 supply chain – closing the gap between research findings and industry practice." *International Journal of Operations & Production Management*, 40(4), pp. 341-355.

Sauer, P.C. and S. Seuring . 2019. "Extending the reach of multi-tier sustainable supply chain management – Insights from mineral supply chains." *International Journal of Production Economics* 217: 31-43.

Schauer S., S. Rass, S. König, T. Grafenauer and M. Latzenhofer. 2018. "Analyzing Cascading Effects among Critical Infrastructures: The CERBERUS Approach. In Proceedings of ISCRAM 2018, Computer Science.

Schauer, S. 2018. "A Risk Management Approach for Highly Interconnected Networks". In: Game Theory for Security and Risk Management. Ed. by S. Rass and S. Schauer. Springer International Publishing: 285–311.

Scott, L.R. 2011. *Numerical Analysis*. Princeton University Press.

SEC Reporting, Release No. 33-7620, (Jan. 5, 1999), URL: https://www.sec.gov/rules/final/33-7620.txt [Last accessed 12/13/2020]

Shekarian, M., S. V. R. Nooraie and M. M. Parast. 2020. "An examination of the impact of flexibility and agility on mitigating supply." *International Journal of Production Economics* 220: 1-16.

Sičanica, Z. and I. Vujaklija. 2020. "Resilience to cascading failures. a complex network approach for analysing the Croatian power grid." *43rd International Convention on Information, Communication and Electronic Technology (MIPRO)*, Opatija, Croatia, 2020; 918-922, doi: 10.23919/MIPRO48935.2020.9245160.

Shi, J., Liu, X., Li, Y., Yu, C. and Han, Y. 2022). "Does supply chain network centrality affect stock price crash risk? Evidence from Chinese listed manufacturing companies." *International Review of Financial Analysis*, 80, pp. 102040, https://doi.org/10.1016/j.irfa.2022.102040

Silverberg G. and B. Verspagen. 2003. "Brewing the future: Stylized facts about innovation and their confrontation with a percolation model, EMAEE Conference, Augsburg, April 10-12, 2003.

Smith A.M., M. Posfai, M. Rohden, A. D. Gonzalez, L. Duenas-Osorio, and R. M. D'Souza. 2019. "Competitive percolation strategies, Scientific Reports, 9:11843, https://doi.org/10.1038/s41598-019-48036-0

Sindhwani, R., J. Jayaram and V. Saddikuti. 2022. "Ripple effect mitigation capabilities of a hub and spoke distribution network: an empirical analysis of pharmaceutical supply chains in India. " ,*International Journal of Production Research*, doi: https://doi.org/10.1080/00207543.2022.2098073

Sokat, K.Y. and Altay N. (2022). "Impact of modern slavery allegations on operating performance." *Supply Chain Management*, https://doi.org/10.1108/SCM-08-2021-0387

Stellinger, A., I. Berglund and H. Isakson. 2020. "How trade can fight the pandemic and contribute to global health.", in (eds.) Baldwin R. E. and Evenett S. J. (2020). COVID-19 and Trade Policy: Why Turning Inward Won't Work, CEPR Press, London, UK.

Sun, J., J. Tang, W. Fu, Z. Chen and Y. Niu. 2020. "Construction of a multi-echelon supply chain complex network evolution model and robustness analysis of cascading failure." *Computers & Industrial Engineering*, 144: 1-16.

Sunmola, F.T. 2021. "Context-Aware Blockchain-Based Sustainable Supply Chain Visibility Management, Procedia Computer Science 180 (2021) 887–892.

Tang L., K. Jing, J. He, and H. E. Stanley. 2016. "Complex Interdependent Supply Chain Networks: Cascading Failure and Robustness." *Physica* (Amsterdam) 443A, 58.

Trajanovski, S., S. Scellato S. and I. eontiadis 2012. "Error and attack vulnerability of temporal networks." PHYSICAL REVIEW E Stat Nonlinear, Soft Matter Phys 85(6):1-10.

Trkman P. and K. McCormack. 2009. "Supply Chain Risk in Turbulent Environments—A Conceptual Model for Managing Supply Chain Network Risk." *International Journal of Production Economics*, 119(2): 247-258.

Viljoen N. M. and J. W. Joubert. 2018. "The Road most Travelled: The Impact of Urban Road Infrastructure on Supply Chain Network Vulnerability." *Networks and Spatial Economics* 18: 85-113.

Wang S., X. Gu, S. Luan and M. Zhao. 2021. "Resilience analysis of interdependent critical infrastructure systems considering deep learning and network theory. International Journal of Critical Infrastructure Protection, 35, 100459.

Wang, H., T. Gu, M. Jin, R. Zhao and H. Wang. 2018. "The complexity measurement and evolution analysis of supply chain network under disruption risks." *Chaos, Solitons and Fractals* 116(1): 72-78.

Wang, S., L. Hong, M. Ouyang, J. Zhang and X. Chen. 2013. "Vulnerability analysis of interdependent infrastructure systems under edge attack strategies." *Safety Science* 51(1): 328-337.

Wierman, J. 2011. "Percolation Theory. Encyclopedia of Statistical Sciences". URL: https://onlinelibrary.wiley.com/doi/pdf/10.1002/9781118445112.stat02317 [Last accessed 09/24/2021]

Williams, M. J. and M. Musolesi. 2016. "Spatio-temporal networks: reachability, centrality and robustness." *Royal Society Open Science* 3(6): 160196.

Wu, T., J. Blackhurst, P. O'Grady. 2007. "Methodology for supply chain disruption analysis." *International Journal of Production Research* 45(7): 1665-1682.

Yacoub, R. and M. El-Zomor. 2020. "Would COVID-19 Be the Turning Point in History for the Globalization Era? The Short-Term and Long-Term Impact of COVID-19 on Globalization" (April 6, 2020. Available at SSRN: https://ssrn.com/abstract=3570142 or http://dx.doi.org/10.2139/ssrn.3570142

Yamashita, K., Y. Yasuda, R. Nakamura and H. Ohsaki. 2019. "Revisiting the Robustness of Complex Networks against Random Node Removal." *Journal of Information Processing* 27: 643-649.

Yan, J., H. He, and Y. Sun. 2014. "Integrated Security Analysis on Cascading Failure in Complex Networks." *IEEE Transactions on Information Forensics and* Security 9(3): 451-463.

Yang, Q., C. M. Scoglio and D. M. Gruenbacher. 2021. "Robustness of supply chain networks against underload cascading failures." *Physica A Statistical Mechanics and its Applications* 563(C): 1-12.

Zegordi, S.H. and H. Davarzani. 2012. "Developing a supply chain disruption analysis model: Application of colored Petri-nets." *Expert Systems with Applications* 39(2): 2102-2111.

Zhao, C., N. Li, and D. Fang, 2015. A Conceptual Framework for Modeling Critical Infrastructure Interdependency: Using a Multilayer Directed Network Model and Targeted Attack-Based Resilience Analysis. In Eds. W.J. O'Brien and Simone Ponticelli, "Computing in Civil Engineering 2015", American Society of Civil Engineers, Reston, VA.

Zheng, K., Y. Liu, Y. Wang, and W. Wang 2021. *"*k-core percolation on interdependent and interconnected multiplex networks." 2021 EPL 133 48003. arXiv:2101.02335

Zhou, Y. and J. Wang. 2018. "Efficiency of complex networks under failures and attacks: A percolation approach.", *Physica A: Statistical Mechanics and its Application* 512(15): 658-664.

Zsidisin, G.A., Ragatz G.L., and Melnyk S. A. (2003). "Effective Practices in Business Continuity Planning for Purchasing and Supply Management." Department of Marketing and Supply Chain Management, The Eli Broad Graduate School of Management, Michigan State University, East Lansing, Michigan. URL: http://citeseerx.ist.psu.edu/viewdoc/download?doi=10.1.1.133.2413&rep=rep1&type=pdf [Last accessed, 3/10/2022]

**Appendix 1**
Supply Chain (SC) Ability Facets: Managing Disruptions

| Supply Chain Abilities | Meaning | Aim | Source |
|---|---|---|---|
| Resiliency | SC's ability to "to recover their performance after having absorbed the disruption effects" and "return to its original [or desired] state after being disturbed." | Recovery | Baz and Ruel (2021); Peck (2003) |
| Robustness | "S.C.s' ability to maintain its planned performance following…disruption(s)." | Maintain operation | Baz and Ruel (2021) |
| Agility | SC's ability to "rapidly align the network and its operations to the dynamic and turbulent requirements of the demand network" and "shifts in supply". | Rapid response | Ismail and Sharifi (2006); Kitchen and Hult (2007) |
| Vulnerability | SC's "exposure to serious disturbance, arising from risks within the supply chain as well as risks external to the supply chain." | Measure of risk | Peck (2003) |
| Flexibility | SC's ability "to respond to changes in the volatile environment, without excessive performance loses." | Manage minor, short-term disruption | Delic and Eyers (2020). |
| Adaptability | SC's ability to "reshape" and "adapt to [an] uncertain environment in order to reduce any adverse impacts…without ties or legacy issues or regard to how the chain has been operated previously." | Manage major, long-term disruption | Chan and Chan (2010); Kitchen and Hult (2007) |

**Appendix 2**.

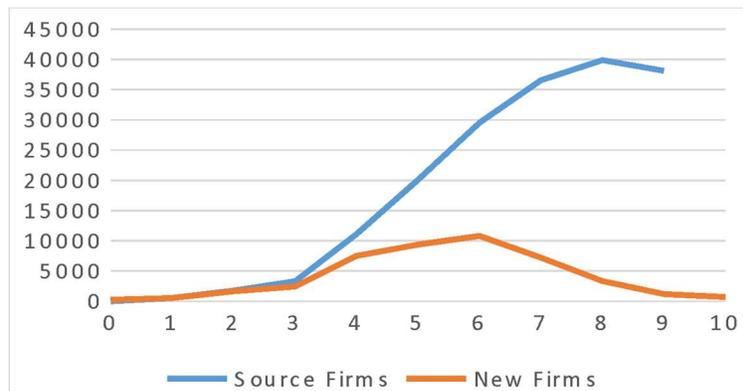

Appendix 2 Caption: Firms' distribution across tiers

Appendix 2 Alt Text: Graph displaying the distribution of source firms and new firms across different tiers.